\documentclass[proof]{WileyASNA-v1}

\articletype{Article Type}%

\received{XXX}
\revised{XXX}
\accepted{XXX}

\newcolumntype{R}{>{$}r<{$}}
\newcolumntype{L}{>{$}l<{$}}
\newcolumntype{A}{R@{${}\pm{}$}L}
\newcolumntype{E}{R@{${}-{}$}L}
\newcolumntype{B}{R@{${}\,/\,{}$}L}
\newcommand{\mcl}[1]{\multicolumn{2}{c}{#1}}
\newcommand{\mcc}[1]{\multicolumn{1}{c}{#1}}

\begin{document}

\title{Astrophysical Parameters and Dynamical Evolution of Open Clusters: NGC~2587, Col~268, Mel~72, Pismis~7}

\author[1]{Hikmet \c{C}akmak*}
\author[2]{Orhan G\"une\c{s}}
\author[1]{Y\"uksel Karata\c{s}}
\author[3]{Charles Bonatto}

\authormark{\c{C}akmak \textsc{et al}}

\address[1]{\orgdiv{Department of Astronomy and Space Sciences}, \orgname{Istanbul University, Science Faculty}, \orgaddress{\state{34119, \"Universite-Istanbul}, \country{Turkey}}}
\address[2]{\orgdiv{Istanbul Medeniyet University, Faculty of Arts and Humanities}, \orgname{Department of History of Science}, \orgaddress{\state{Kadikoy, Istanbul}, \country{Turkey}}}
\address[3]{\orgdiv{Universidade Federal do Rio Grande do Sul}, \orgname{Departamento de Astronomia}, \orgaddress{\state{CP\,15051, RS, Porto Alegre 91501-970}, \country{Brazil}}}

\corres{*Hikmet \c{C}akmak \email{hcakmak@istanbul.edu.tr}}

\keywords{open clusters and associations: individual NGC 2587, open clusters and associations: 
individual Col 268, open clusters and associations: individual Mel 72, open clusters and associations: individual Pismis 7}

\abstract{We determined astrophysical and dynamical parameters of the open clusters (OCs) NGC\,2587, Collinder\,268 (Col\,268), Melotte\,72 (Mel\,72), and Pismis\,7 from Gaia DR2 photometric/astrometric data and a new technique, \textit{fitCMD}.  

\textit{fitCMD} provides (Z,~Age(Gyr)) as (0.025,~0.45) for NGC~2587, (0.0025,~0.5) for Col.~268, (0.011,~1.25) for Mel~72, and (0.008,~1.00) for Pismis~7, respectively. As compared to Gaia DR2 distances , the obtained photometric distances from \textit{fitCMD} provide somewhat close distances. For NGC\,2587 and  Mel.\,72, both distances are in good concordance. 

Except for NGC\,2587, the ages of the remaining OCs are higher than their relaxation times, which suggests that they are dynamically relaxed. NGC\,2587 did not undergo dynamical evolution. Mel\,72 and Pismis\,7 with relatively flat MF slopes indicate signs of a somewhat advanced dynamical evolution, in the sense that they appear to have lost a significant fraction of their low-mass stars to the field.
Pismis\,7's negative/flat MFs indicates that its high mass stars slightly outnumber its low mass ones. Given its mild dynamical evolution, the high mass stars move towards the central region, while low-mass stars are continually being lost to the field. Col\,268 presents small dimensions which suggest a primordial origin. The outer parts  of Mel\,72 and Pismis\,7 - with large cluster radii expand with time, while Mel\,72's core contracts because of dynamical relaxation (Figs.~10e--f).}

\jnlcitation{\cname{%
\author{\c{C}akmak H.}, 
\author{G\"une\c{s} O.}, \author{Karata\c{s} Y.}, and
\author{Charles Bonatto}} (\cyear{XXXX}), 
\ctitle{Revealing Dynamical Properties of Open Clusters: NGC~2587, Col~268, Mel~72, Pismis~7}, \cjournal{AN}, \cvol{XXX;XX:X--X}.}

\maketitle

\section {Introduction}
Astrophysical parameters (reddening, distance, age), structural parameters (core and cluster radii, $R_{core}$ and $R_{RDP}$), and overall masses, mass function slopes ($\chi$, MF slopes), relaxation times $(t_{rlx})$ and evolutionary parameters $(\tau)$ of the open clusters (OCs) are needed for the interpretation of the dynamical evolution. The stars inside the OCs undergo internal and external perturbations such as stellar evolution, mass segregation, and encounters with the disk and Giant Molecular Clouds (GMCs) \citep{Lamers2006}, \citep{Gieles2007}. All these process produce a varying degree of mass loss that may lead to the cluster dissolution into the field. Mass segregation preferentially drives the low mass stars to the outer parts of the clusters.

Our sample OCs, NGC\,2587, Col.\,268, Mel.\,72, and Pismis\,7 were studied in 2MASS~$JHK_{s}$ by \citet[hereafter, B11]{Bukowiecki2011} and \citet[hereafter, K13]{Kharchenko2013}. \cite{Tadross2011} published the astrophysical parameters of NGC~2587 in 2MASS. The astrophysical parameters of Mel\,72 were derived in $UBVI$ photometry by \cite{Hasegawa2008} and  \cite{Piatti2009}. B11 published their astrophysical/structural parameters, and their masses/MF slopes, except for Pismis\,7. K13 also gave their astrophysical parameters together with  their core radii (Tables 6--7). 

The likely members of four OCs have been determined from Gaia DR2 astrometric/photometric data \citep{bro18,lin18}. To study the issues above, the well-determined cluster parameters (heavy element abundance, colour excess, distance modulus/distance, age and mass) are needed. For this, we employ an approach \textit{fitCMD}, designed by \cite{Bonatto2019}. \textit{fitCMD} uses the isochrones of \cite{bre12} (hereafter B12) and $G$, $G_{BP}$,~ $G_{RP}$ filters.

This paper is organized as follows. The cluster membership technique is presented in Section~2. The derivation of reddenings, distance moduli/distances, ages of four OCs from \textit{fitCMD} algorithm is explained in Section~3. The obtained cluster dimensions, masses/mass function slopes, relaxation times/ evolutionary parameters of four OCs are given in Sections~4--5. A Discussion/Conclusion is presented in Section~6 together with the sub-sections; a comparison with the literature and the dynamical evolution. 

\renewcommand{\tabcolsep}{2.4mm}
\renewcommand{\arraystretch}{1.1}
\begin{table}[h!]\label{Table-1}
	\begin{center}
		\caption{Equatorial and Galactic coordinates of four OCs.}
		\begin{tabular}{lrrrr}
			\hline
			Cluster   &$\alpha(2000)$&$\delta(2000)$&$\ell$&$b$\\
			          &(h\,m\,s)     &$(^{\circ}\,^{\prime}\,^{\prime\prime})$&$(^{\circ})$&$(^{\circ})$ \\
			\hline
			NGC\,2587 &  08 23 22.9  & -29 31 02.1  &  249.46  &  4.46\\
			Col\,268  &  13 18 11.4  & -67 05 00.0  &  305.54  & -4.35\\
			Mel\,72   &  07 38 31.3  & -10 41 51.4  &  227.84  &  5.38\\
			Pismis\,7 &  08 41 08.0  & -38 42 08.6  &  259.05  &  1.99\\
			\hline
		\end{tabular}
	\end{center}
\end{table} 

\renewcommand{\tabcolsep}{1.3mm}
\renewcommand{\arraystretch}{1.6}
\begin{table}[b!]\label{table-2}
	\centering
	{\footnotesize
		\caption{The median proper motion components, proper motion radii and parallaxes/distances of the likely members of four OCs for this paper (top rows) and \cite{cantat2020} (bottom rows).}
		\begin{tabular}{lccccc}
			\hline
			Cluster & $\mu_{\alpha}$ (mas/yr) & $\mu_{\delta}$ (mas/yr)& $\Delta r$ (mas/yr) &$\varpi$ (mas) & $d(kpc)$  \\  
			\hline
			NGC\,2587  & -4.27$\pm$0.08  &  3.57$\pm$0.08 & 0.21 & 0.31$\pm$0.03 & 3.25$\pm$0.08 \\
			Col\,268   & -5.48$\pm$0.08  & -0.40$\pm$0.07 & 0.19 & 0.37$\pm$0.06 & 2.67$\pm$0.13 \\
			Mel\,72    & -4.18$\pm$0.08  &  3.66$\pm$0.07 & 0.21 & 0.38$\pm$0.05 & 2.61$\pm$0.08 \\
			Pismis\,7  & -3.31$\pm$0.09  &  2.77$\pm$0.10 & 0.15 & 0.19$\pm$0.04 & 4.82$\pm$0.35 \\
			\hline
			NGC\,2587  & -4.26$\pm$0.09  &  3.59$\pm$0.11 &      & 0.30$\pm$0.05 & 3.39$\pm$0.59 \\
			Col\,268   & -5.48$\pm$0.09  & -0.40$\pm$0.08 &      & 0.36$\pm$0.04 & 2.80$\pm$0.30 \\
			Mel\,72    & -4.16$\pm$0.11  &  3.68$\pm$0.09 &      & 0.37$\pm$0.06 & 2.71$\pm$0.40 \\	
			Pismis\,7  & -3.31$\pm$0.09  &  2.79$\pm$0.13 &      & 0.15$\pm$0.06 & 6.67$\pm$2.44 \\
			\hline  
			
		\end{tabular}
	}
\end{table}

\begin{figure*}[t!]\label{Fig-1}
	\centering{
		\includegraphics[width=0.38\textwidth]{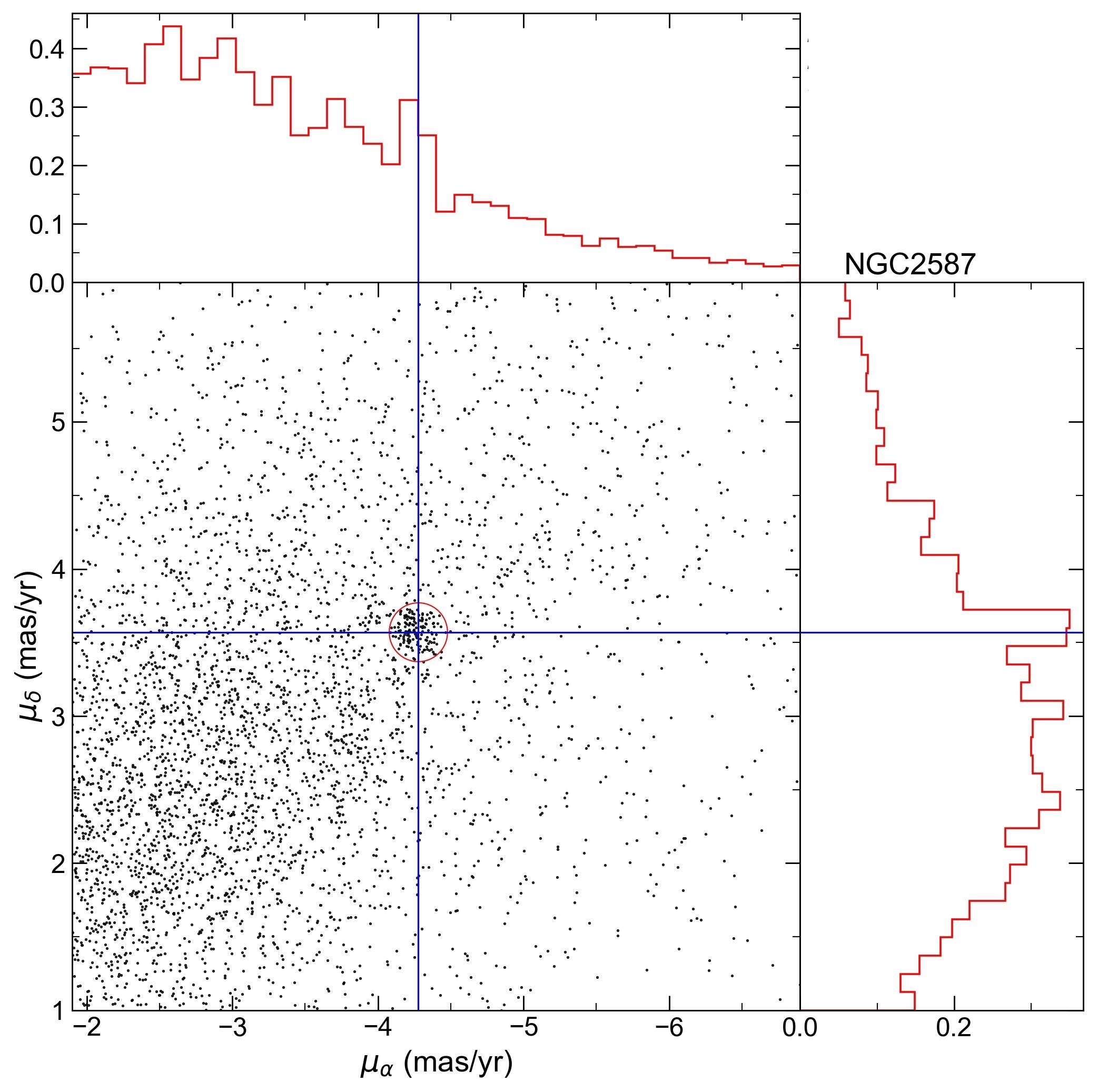}\hspace{5mm}
		\includegraphics[width=0.38\textwidth]{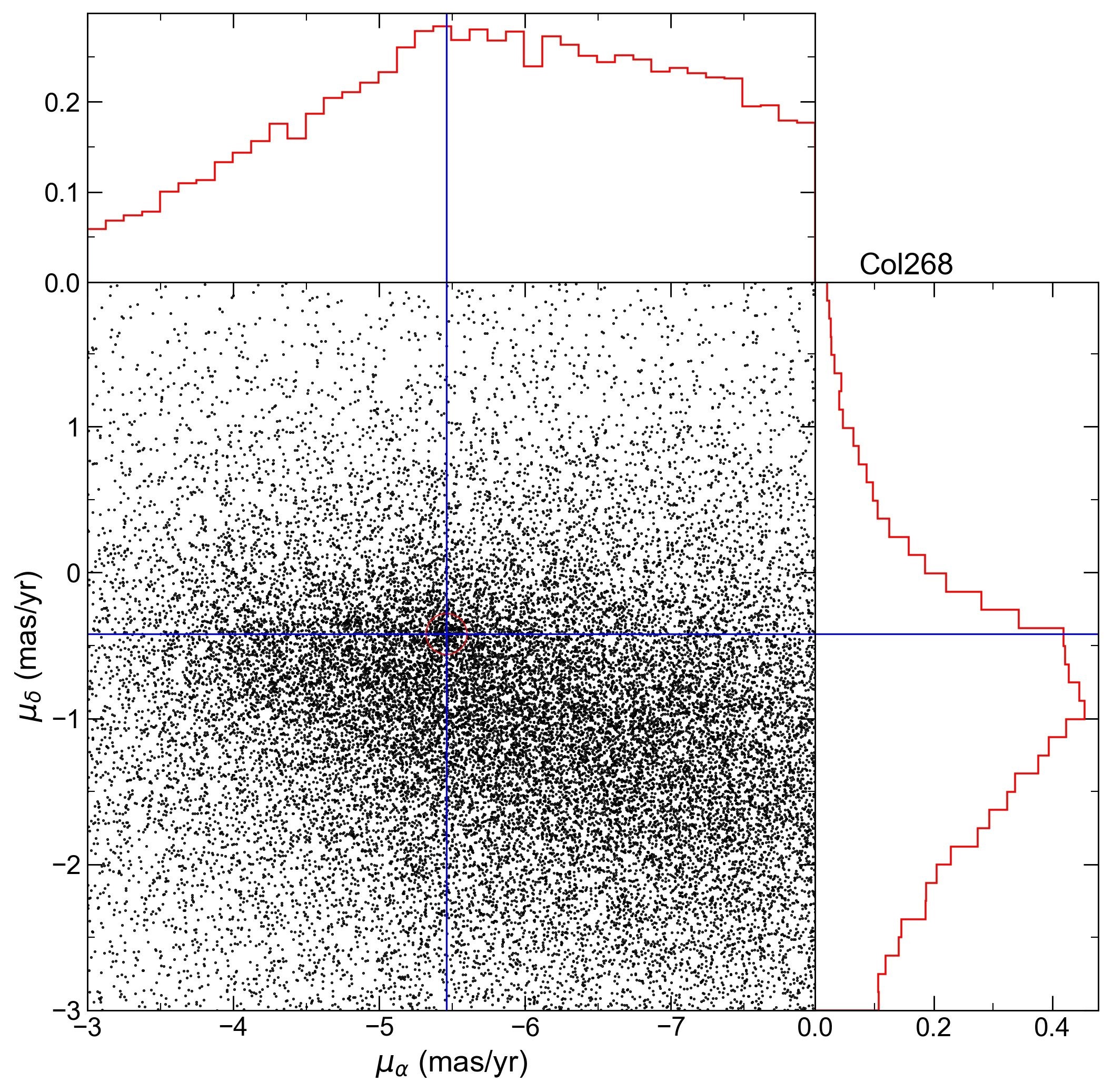}\\[2ex]
		\includegraphics[width=0.36\textwidth]{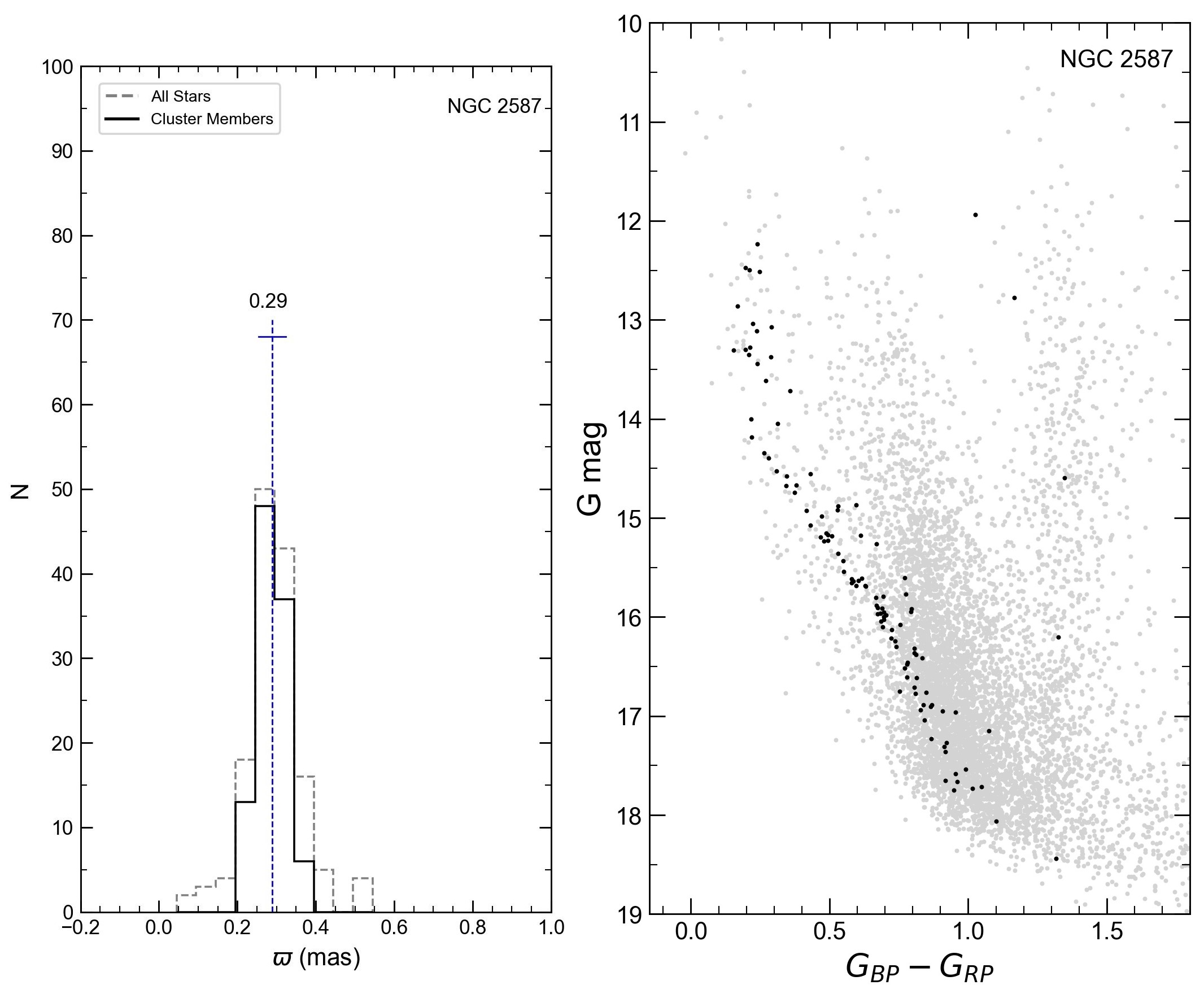}\hspace{22mm}
		\includegraphics[width=0.36\textwidth]{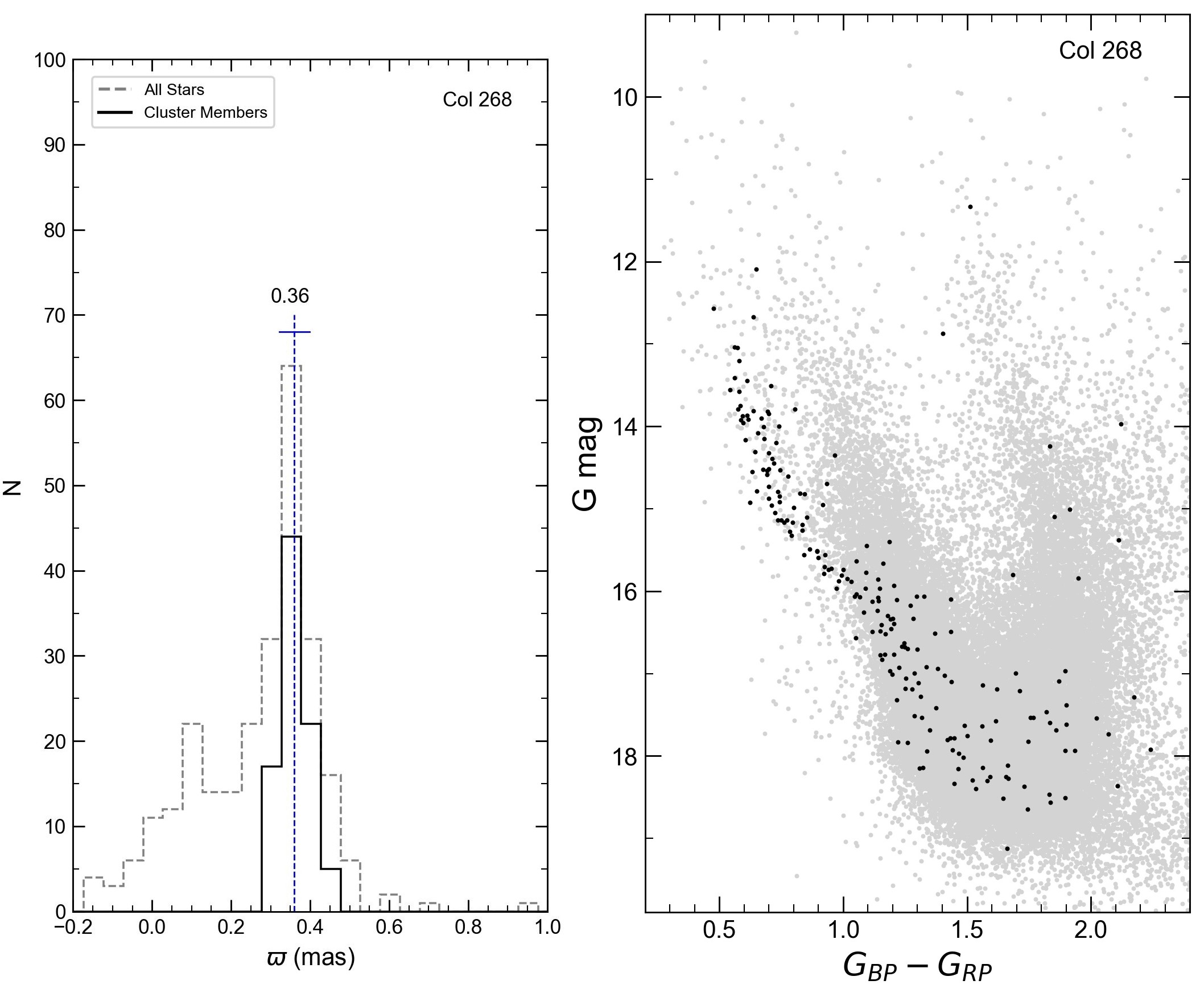}
	}
	\caption{The $\mu_{\alpha}$ versus $\mu_{\delta}$, parallax histogram plus Gaia CMD for NGC 2587 and Col 268. The red circles indicate the cluster region. The parallax histograms of the potential members inside the red circles are shown in the bottom left. The cluster stars inside $\varpi \pm \sigma_{\varpi}$~mas (the vertical/horizontal lines on the parallax histograms) are our likely members. A single stellar cluster sequence of the probable members (filled dots of bottom right panels) is clearly seen on the $(G, G_{BP}-G_{RP})$ CMDs.}
\end{figure*}

\begin{figure*}[t!]\label{Fig-2}
	\centering{
		\includegraphics[width=0.38\textwidth]{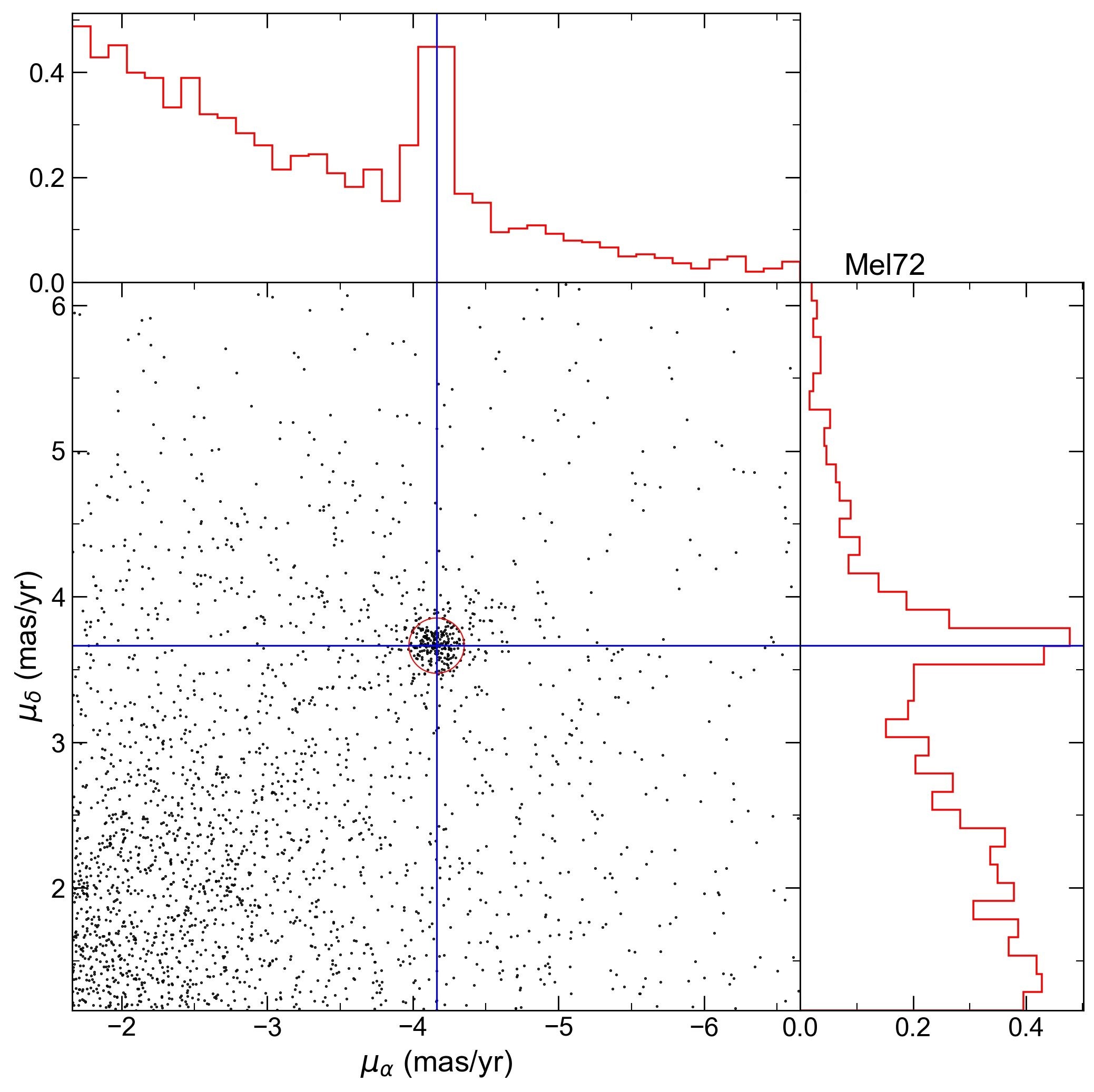}\hspace{5mm}
		\includegraphics[width=0.38\textwidth]{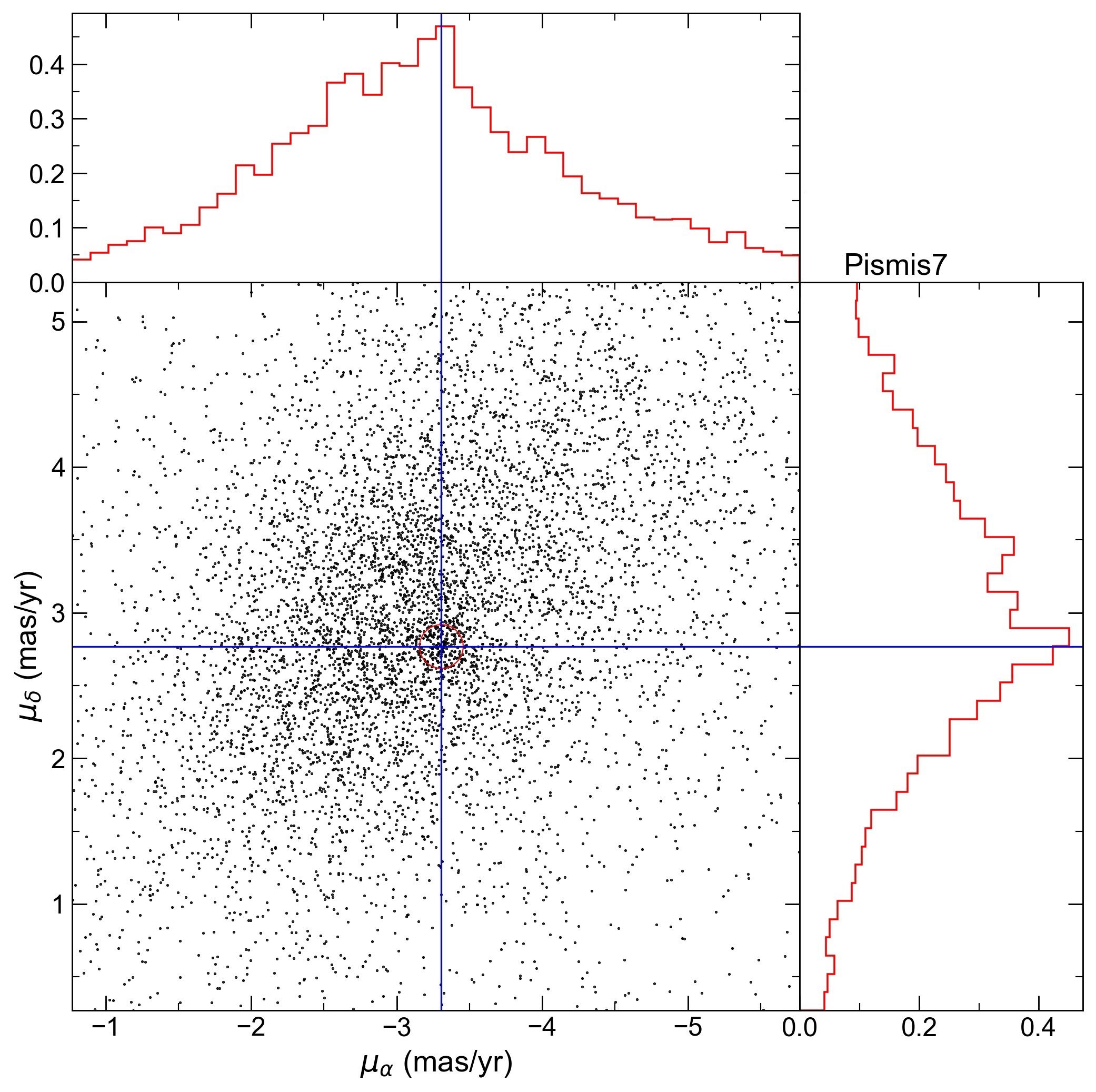}\\[2ex]
		\includegraphics[width=0.36\textwidth]{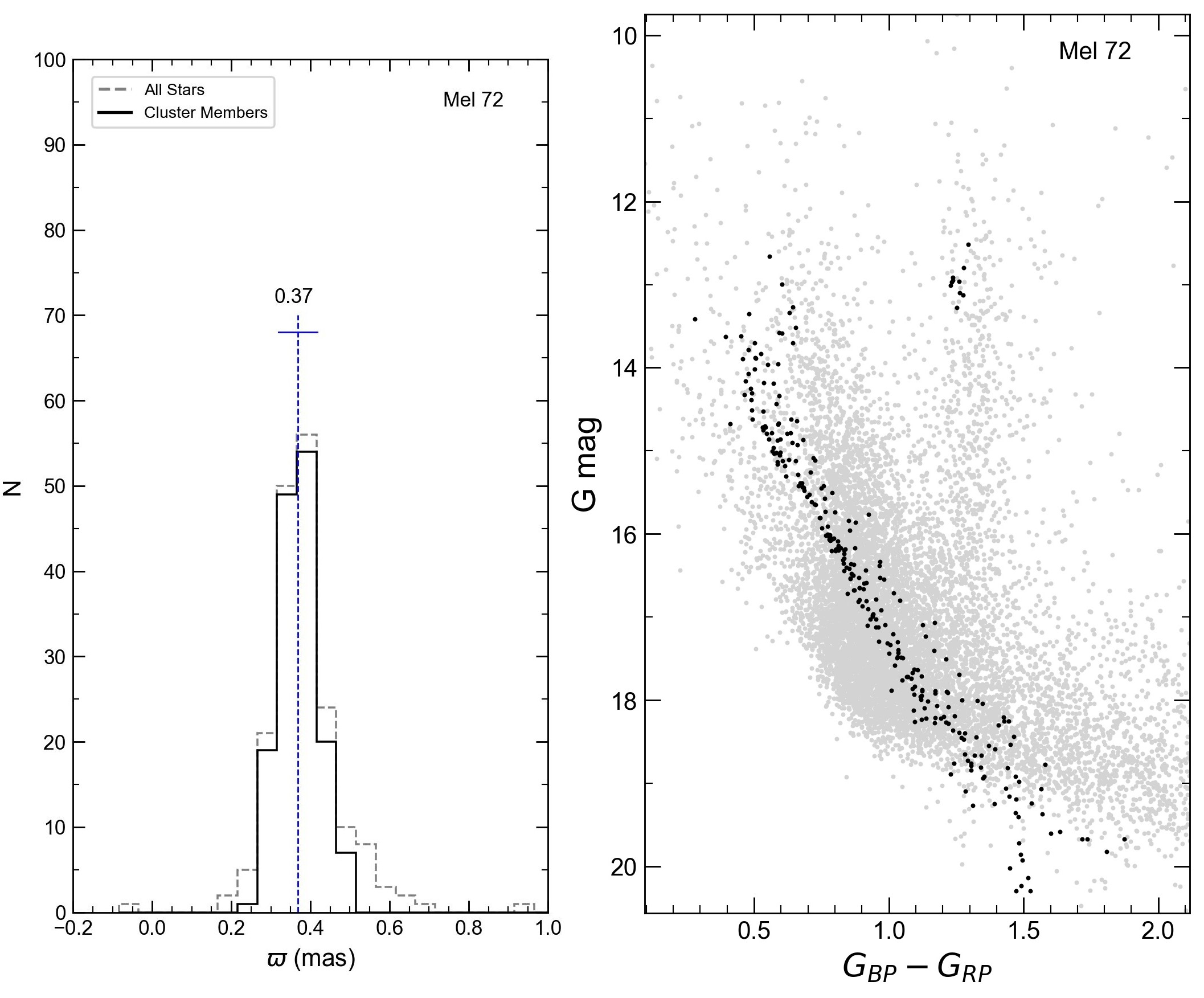}\hspace{22mm}
		\includegraphics[width=0.36\textwidth]{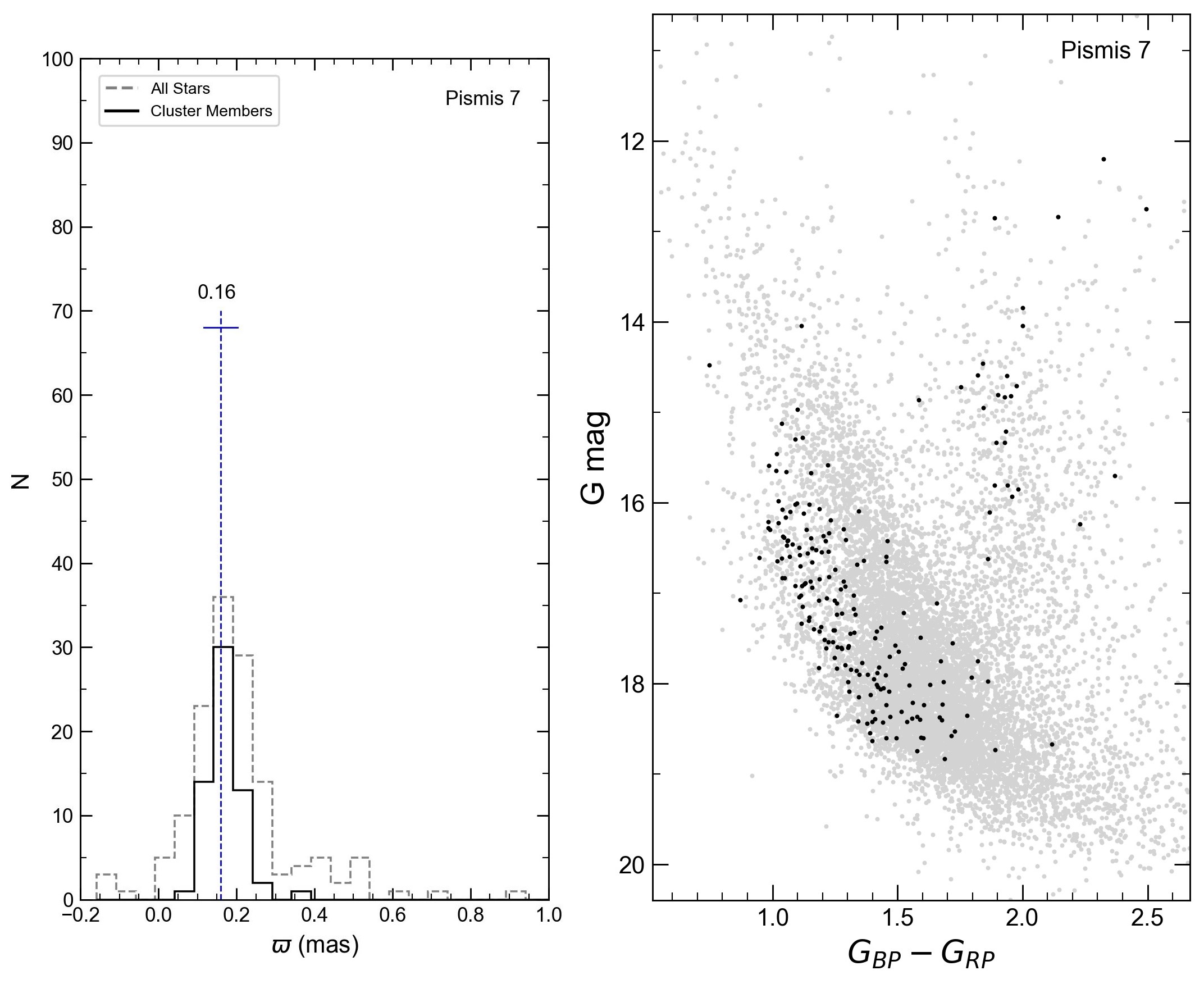}
	}
	\caption{The $\mu_{\alpha}$ versus $\mu_{\delta}$, parallax histogram plus Gaia CMD for Mel 72 and Pismis 7. The symbols and their meanings are the same as Fig.1.}
\end{figure*}

\section{Separation of the Cluster Members}
In order to separate the cluster members of NGC~2587, Col~268, Mel~72 and Pismis~7, we have obtained their Gaia DR2 astrometric/photometric data \citep{bro18} for a large-area (15 arcmin) from VizieR\footnote{http://vizier.u-strasbg.fr/viz-bin/VizieR?-source=II/246.}. 
Their cluster stars are shown on the ($\mu_{\alpha}$,~$\mu_{\delta}$) diagrams (hereafter, VPD) (Figs.~1--2). The potential cluster members show a more concentrated structure, whereas field stars have a more scattered distribution. The eye-fitted proper motion radii (red circles)  represent the cluster region. They have been constructed via the mathematical equations, $x = x_{0} + r\cos(\theta)$ and $y = y_{0} + r\sin(\theta)$. Here, (x$_{0}$, y$_{0}$) are the median values of ($\mu_{\alpha}$, $\mu_{\delta}$) (mas~yr$^{-1}$), the radius $r = \sqrt{\sigma_{\mu\alpha}^{2} + \sigma_{\mu\delta}^{2}}$ mas~yr$^{-1}$, and $\theta$ =$0^{\circ}$ to $360^{\circ}$. These propor motion radii are listed in Col.~4 of Table~2, and they are a good compromise for the likely cluster members. The median values of the proper motion histograms on top/right of the VPDs (vertical/horizontal blue lines) make it easier to determine the proper motion value of the cluster's center.
The median values ($\varpi \pm \sigma_{\varpi}~mas$) of the parallax histograms (bottom-left panels) for the cluster stars inside the red circles classify the probable members.  By applying Gaussian Mixture Model (GMM) \cite{ped11} to the probable members inside $\varpi \pm \sigma_{\varpi}$~mas,  their membership probabilities, P($\%$) have also been determined in order to check. They have greater than 80$\%$ (vertical dashed lines of Fig.~3). The GMM \footnote{$P$ is defined $\Phi_c$ /$\Phi$.  Here $\Phi = \Phi_c + \Phi_f$ is the total probability distribution. \textit{c} and \textit{f} subscripts for cluster and field parameters, respectively. The used parameters for estimation of $\Phi_c$ and $\Phi_f$ are $\mu_{\alpha}$, $\mu_{\delta}$, $\varpi$, $\sigma_{\mu\alpha}$, $\sigma_{\mu\delta}$, $\sigma_\varpi$.} model considers that the distribution of proper motions of the stars in a cluster's region can be represented by two elliptical bivariate Gaussians. The expressions used can be found in the papers of \cite{bal98}, \cite{wu02}, \cite{sar12} and \cite{dia18}.
These probable members provide good single stellar sequence on the Gaia DR2 $(G, G_{BP}-G_{RP})$ colour–magnitude diagrams (CMDs) (bottom-right panels of Figs.~1--2), and thus they are considered for determining the astrophysical parameters. Note that the scatter of the members on the $(G, G_{BP}-G_{RP})$ of Pismis~7 is similiar to Fig.~2 of \cite  {Ahumada2005}.  
The estimated median equatorial coordinates from the likely members of our OCs are listed in Table 1. These coordinates are similar to those of WEBDA \citep{Mermilliod1992}.

Instead of the inverse of Gaia-DR2 parallaxes, the distances of four OCs are obtained from the posterior probability density functions (PDFs)\footnote{https://www2.mpia-hd.mpg.de/~calj/gedr3-distances/main.html} given by \cite{Bailer2018} and \cite{Bailer2021}. For this, we use the global zero point of $-$0.017 mas \citep{lin21}.
The median proper motion components/the proper motion radii and the median parallaxes/distances of four OCs are listed in Table 2. Within the errors, their median values are in compatible with those of \cite{cantat2018} and \cite{cantat2020} (bottom panel of Table 2). However, note that both median parallaxes of Pismis~7 are somewhat different.

\section{Derivation of Reddening, Distance, Age and Mass}
In order to obtain the astrophysical parameters from Gaia DR2 $(G, G_{BP}-G_{RP})$ for the probable members of four OCs, an approach (\textit{fitCMD}), improved by \cite{Bonatto2019} is utilised.  \textit{fitCMD} is to transpose theoretical initial mass function $(IMF)$ properties for the isochrones of given age and metallicity  to  their observational CMDs. Based on initial mass function (IMF) properties of the B12 PARSEC isochrones\footnote{http://stev.oapd.inaf.it/cgi-bin/cmd.}, \textit{fitCMD} searches for the values of cluster stellar mass $(m_{cl})$, $Age$, global metallicity ($Z$), foreground reddening $E(G_{BP}-G_{RP})$ , distance modulus $(m-M)_{G}$, and magnitude-dependent photometric completeness that produce the artificial and observational CMDs. The adopted parameter ranges are presented in Table 3. The photometric completeness limits which are estimated from Fermi function are given in row 9 of Table~3.

For the members, a Gaia DR2 CMD with the IMF is built for each isochrone (at $DM=0$ and no reddening). By CMD cell- containing stars with mass in the range $(m1,m2)$, the following parameters are obtained, the number of stars per cluster mass $n_{H}=N_{1,2}/m_{cl}$ and  the average mass ($<m>$), which basically involves integrating the mass function  ($\phi(m) = dN/dm$) between $(m1,m2)$: $N_{1,2} = \int_{m1}^{m2}\phi(m)~dm$. Each CMD cell contains the respective relative density (number per cluster mass) of occurrence of stars, which is equivalent to the classical Hess diagram, $H_{M} = H_{M}(m_{cl}, Age, Z)$.  
The parameter search consists on finding the absolute minimum of the residual hyper-surface $R_{H}$  
defined as  $R_{H}=(H_{obs}-H_{sim})^2$, where $H_{obs}$ is the observed Hess diagram and 
$H_{sim}=H_{sim}(m_{cl}, Age, Z, CE, DM, k_{F}, m_{TO})$ is the simulated one. Here, $CE$, $DM$, $k_{F}$, $m_{TO}$ mean the colour excess, distance modulus, the steepness of the descent, and turn-off magnitude, respectively. The adopted thresholds/Hess cell numbers/Hess widths for Gaia colours/magnitudes are listed in rows 7--8 of Table 3. The values for seven parameters at the absolute minimum are assumed to represent those of the star cluster. In practical terms, locating the minima of $R_{H}$ is equivalent to finding the parameters that minimize the  quantity $S_{R}$. Here, 
$S_{R} = \sum_{mag,col}W(mag)\times\ R_{H}(mag,col)$ on the colour/magnitude plane. The sum runs 
over all Hess cells and $W(mag)$ is the statistical weight of each cell. $W(mag)$ corresponds to the 
inverse of the observed Hess density computed at the respective magnitude of each cell, i.e. 
$W(mag)=1/\sum_{col} H_{obs}(mag,col)$.
Residual minimization between observed and synthetic CMDs – by means of the global optimization algorithm $SA$ – then leads to the best-fitting parameters.
$SA$ (Simulated Annealing) is an iterative and statistical technique. For any given step $k$, it randomly selects a new set of parameters ($M^k_{cl}$, $Age^k$, $Z^k$, $CE^k$, $DM^k$, $k^k_{F}$, $m^k_{TO}$) from the respective  ranges.  thus, $SA$ concentrates on smaller parameter ranges, centred around the most promising values, and a new  step $(k+1)$ is taken. Iterations stop when $SA$ meets the convergence criterion: five consecutive repetitions of the same value of  $S_{R}$. However, for the statistical nature of $SA$, \textit{fitCMD} should be repeated a few times to minimize the probability of getting stuck into a deep, but secondary minimum. 

\renewcommand{\tabcolsep}{2.5mm}
\renewcommand{\arraystretch}{1.3}
\begin{table*}[t!]\label{table-3}
	\centering
	{\footnotesize
		\caption{The parameters which are used for \textit{fitCMD}.}
		\begin{tabular}{lllll}
			\hline
			Cluster &   NGC 2587 &    Col 268 &     Mel 72 &   Pismis 7 \\
			\hline
			Cluster (stellar) mass (m$_{\odot}$) & 100-1000& 10-10000& 100-1000 & 10-10000 \\
			$(m-M)_{G}$~(mag)                    &  10.0-15.0 &  10.8-15.1 &  10.0-15.0 &  12.0-17.0 \\
			$E(G_{BP}-G_{RP})$~(mag)             &  0.00-1.00 & 0.0074-1.76 &  0.00-1.00 &  0.00-2.00 \\
			Age~(Myr)                            & 100 - 1000  &   100-1000 &  100-10000 &  120-13500 \\
			$Z$~$(Z_{\odot}= +0.0152)$             & 0.0001 - 0.03 & 0.0001-0.03 & 0.0001-0.03 & 0.0001-0.03 \\
			$A_{V}$~(mag)                        & 3.1 &        3.1 &        3.1 &        3.1 \\
			$(G_{BP}-G_{RP})$ threshold/Hess Cell/Cell width ~(mag) & (0.08-1.41)/67/0.020  & (0.37-2.20)/92/0.020 & (0.19-1.64)/73/0.020 & (0.63-2.58)/98/0.020 \\
			$G$~threshold/Hess Cell/Cell width~(mag) & (11.75-18.06)/64/0.100 & (11.15-17.58)/65/0.100 & (12.31-19.40)/71/0.101 & (12.00-18.83)/69/0.101 \\
			Photometric completeness~(mag)       &$G < 17.17$  &$G < 16.46$   & $G < 17.02$  & $G < 18.08$ \\
			Usable members                       &   118 &        128 &        273 &        204 \\
			\hline
		\end{tabular}
	}
\end{table*} 

\begin{figure}[h!]\label{Fig-3}
	\centering{\includegraphics[width=0.75\columnwidth]{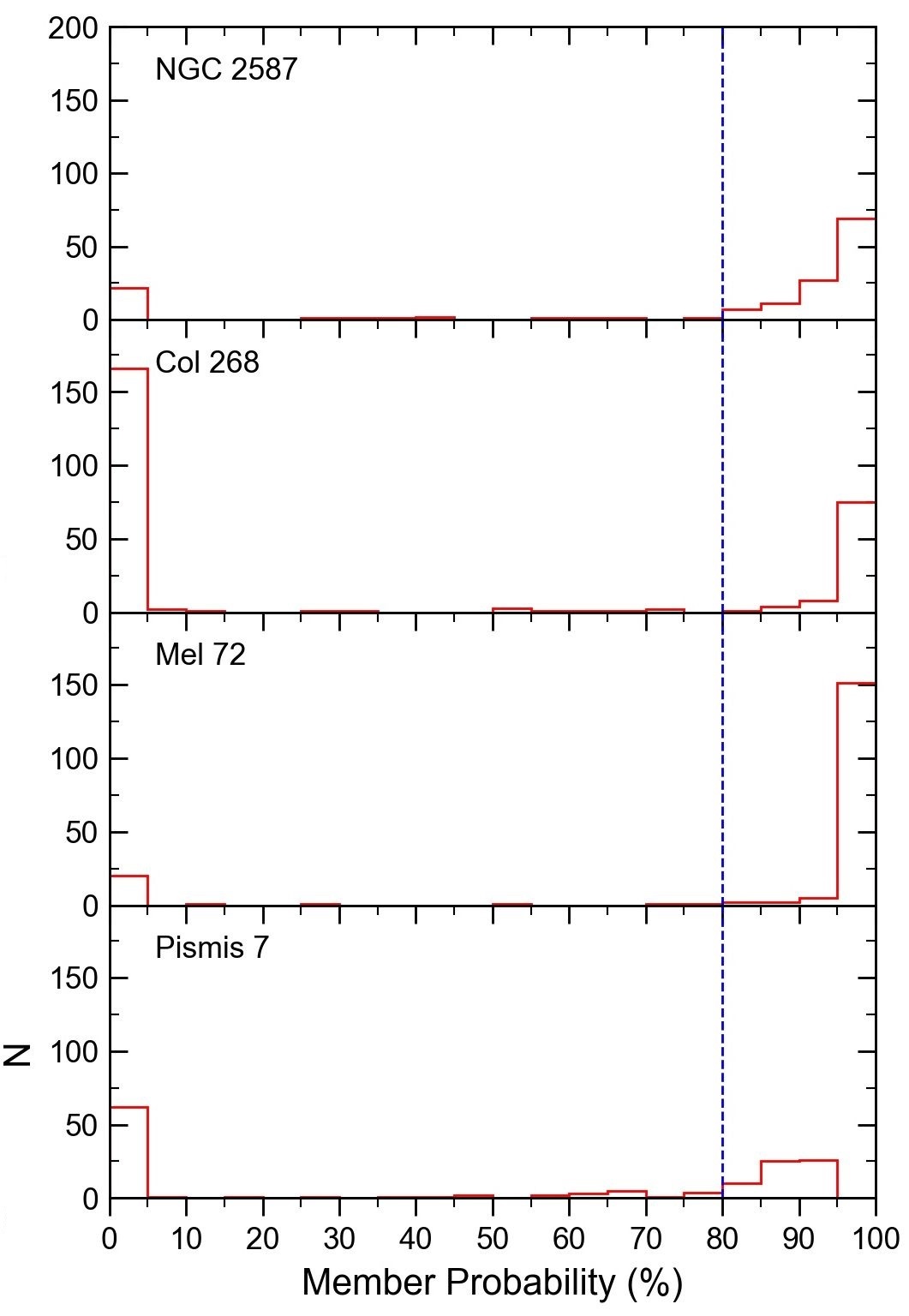}}
	\caption{The membership probability histogram P($\%$) of four OCs. The vertical blue dotted lines} show the likely members which are greater than 80$\%$ inside $\varpi \pm \sigma_{\varpi}$~mas.
\end{figure}

The efficiency of \textit{fitCMD} for input parameters is tested with simulated CMDs, built with pre-defined values of seven parameters. Individual stellar masses are attributed  according to Kroupa's IMF \citep{Kroupa2001} - for masses larger than $0.1~m_{\odot}$ - until the sum  matches $m_{cl}$; the respective magnitudes are taken from the PARSEC isochrone corresponding to $Age$ and $Z$. Typical photometric uncertainties are added for each Gaia DR2 filters and their uncertainties $(\sigma_{k})$.  If the age, metallicity, distance  modulus and colour excess are reasonably well determined, the ratio observed/artificial Hess cells content corresponds to  $m_{cl}$.

The obtained best-fitting astrophysical parameters $E(G_{BP}-G_{RP})$, $E(B-V)$, $Z$, $(V-M_{V})_{0}$, d~(pc) and Age (Gyr) are given in Table~6. The observed $(G, G_{BP}-G_{RP})$ CMDs (Hess diagrams) are shown in Figs.~4--5. The B12 isochrones (solid red lines) fit well the main sequence (MS), turn$-$off (TO) and Red Giant/Red Clump (RG/RC) regions. 

\begin{figure*}[t!]\label{Fig-4}
	\centering{
		\includegraphics[width=0.35\textwidth]{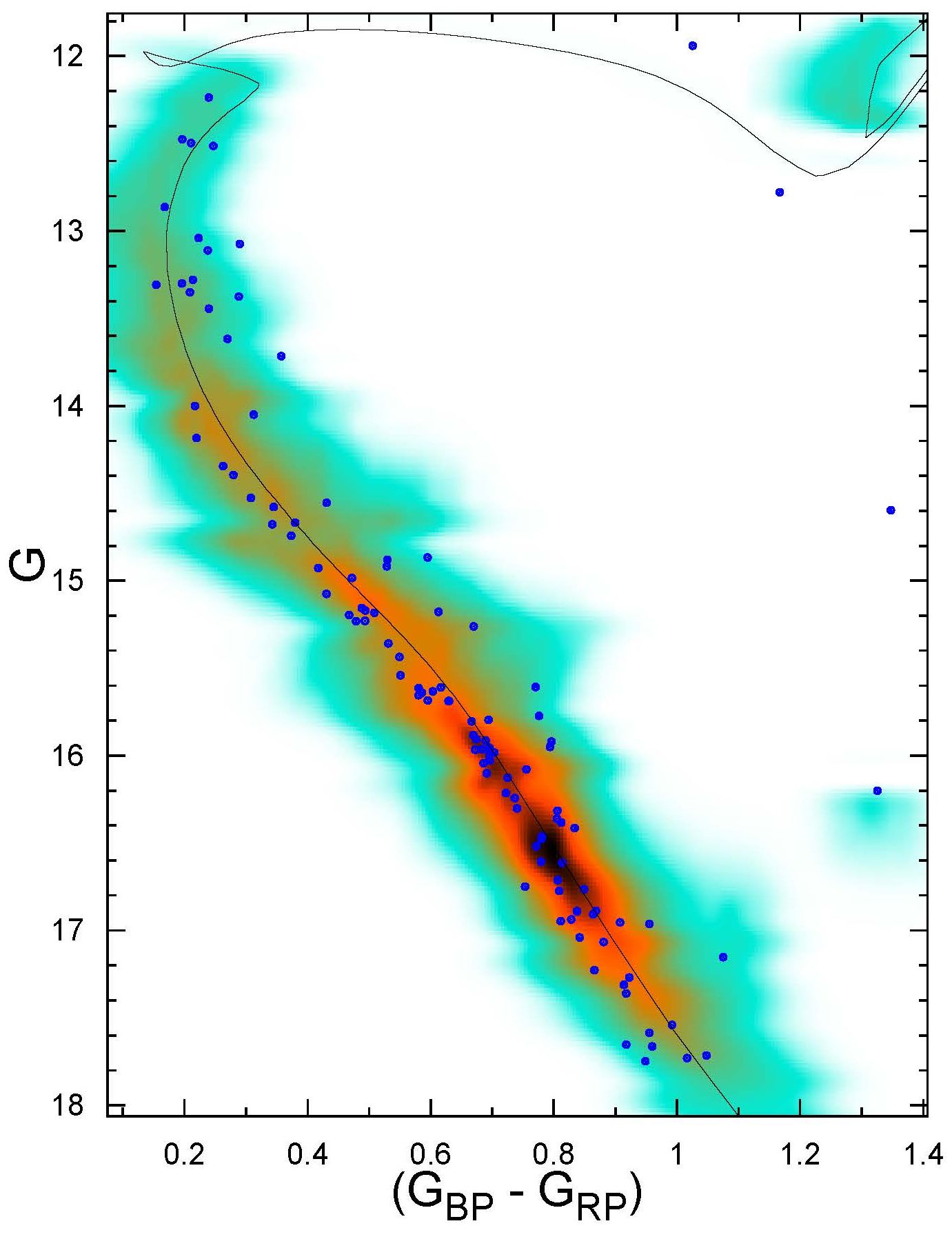}\hspace{8mm}
		\includegraphics[width=0.35\textwidth]{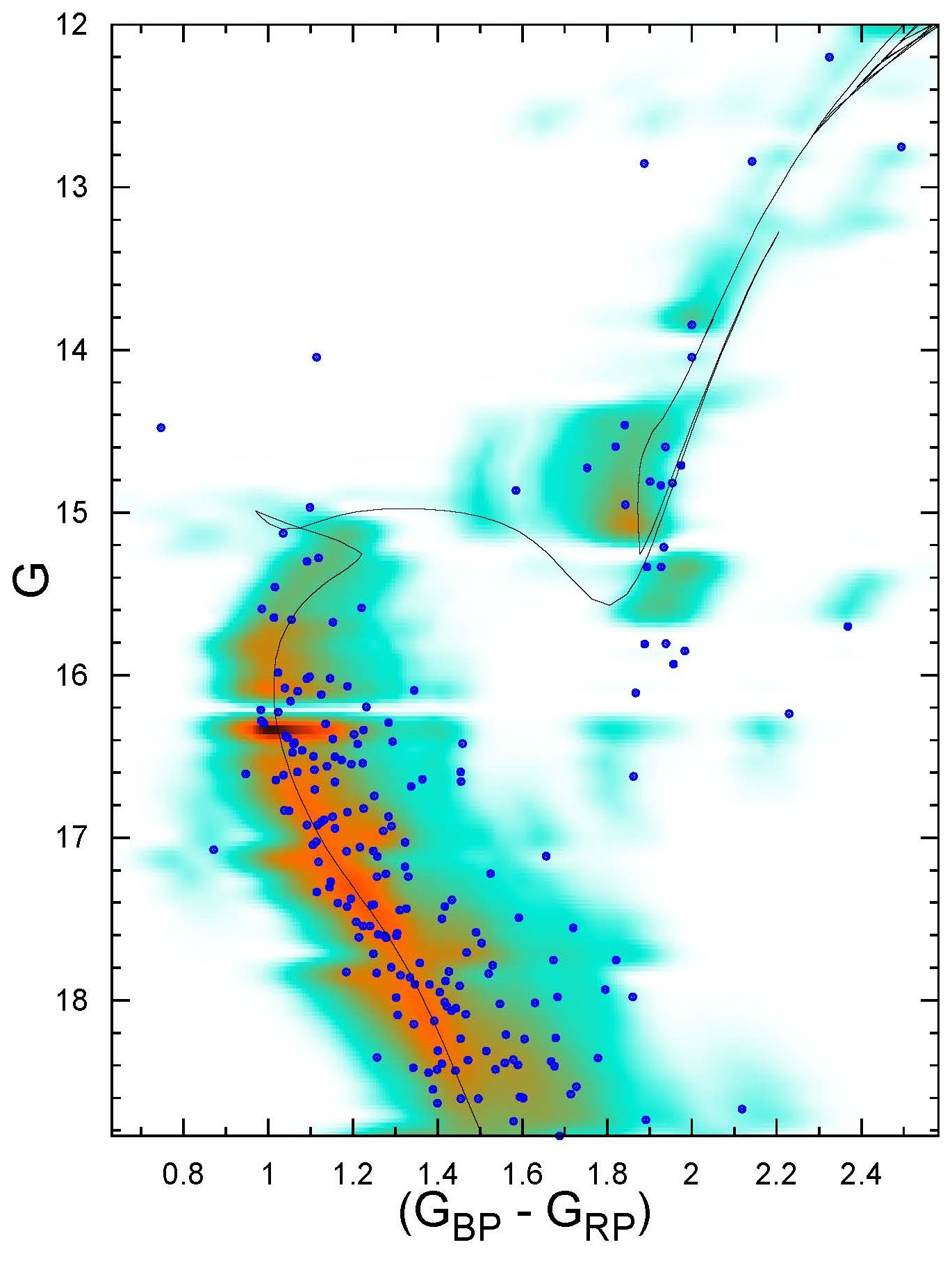}
	}
	\caption{$(G, G_{BP}-G_{RP})$ CMDs (Hess~diagram) for NGC~2587 and Col~268. The blue dots represent the probable members (the bottom rows of Table 3). Solid black lines denote the B12 isochrones. The coloured shaded regions show the relative stellar densities (number per cluster mass) within CMD cells.}
\end{figure*}

\begin{figure*}[t!]\label{Fig-5}
	\centering{
		\includegraphics[width=0.35\textwidth]{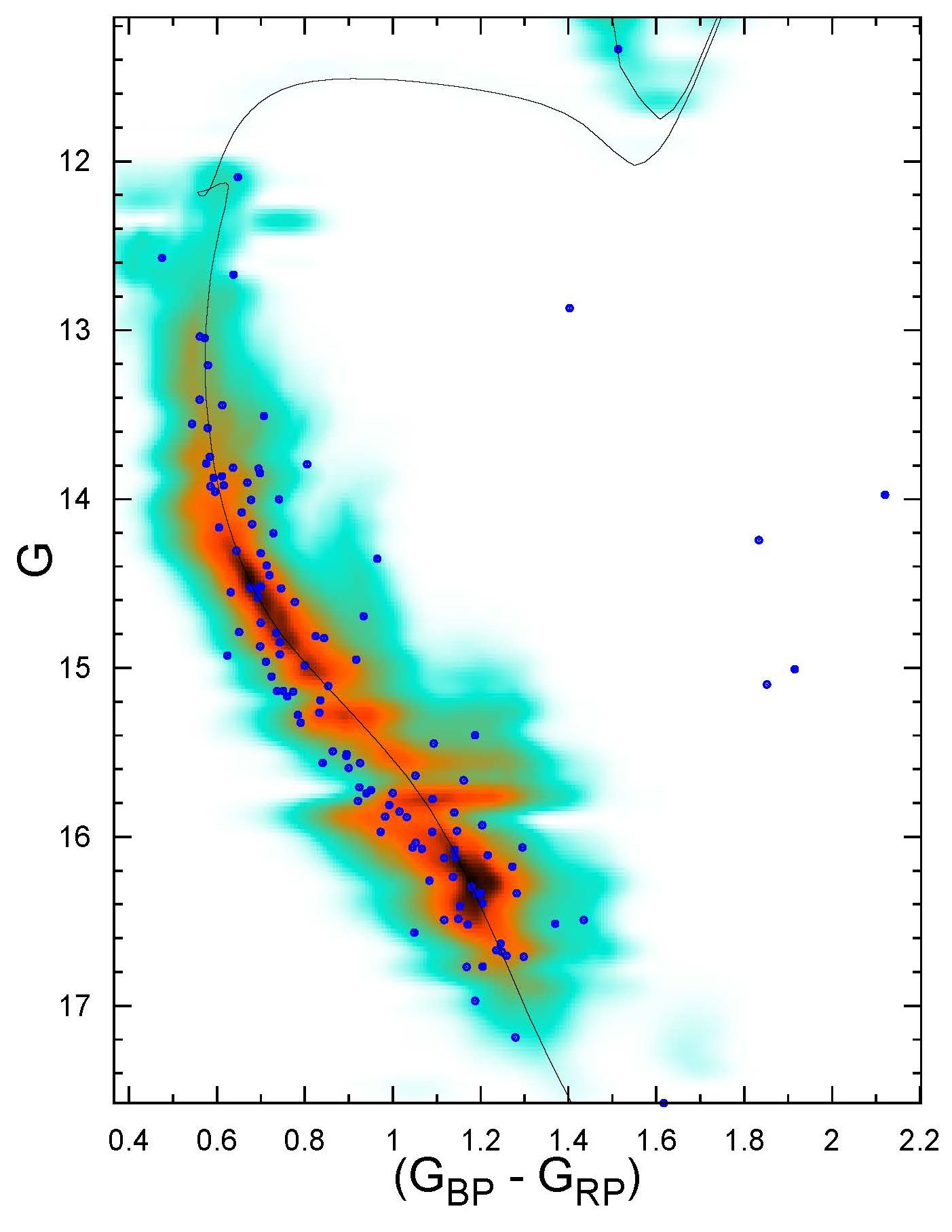}\hspace{8mm}
		\includegraphics[width=0.35\textwidth]{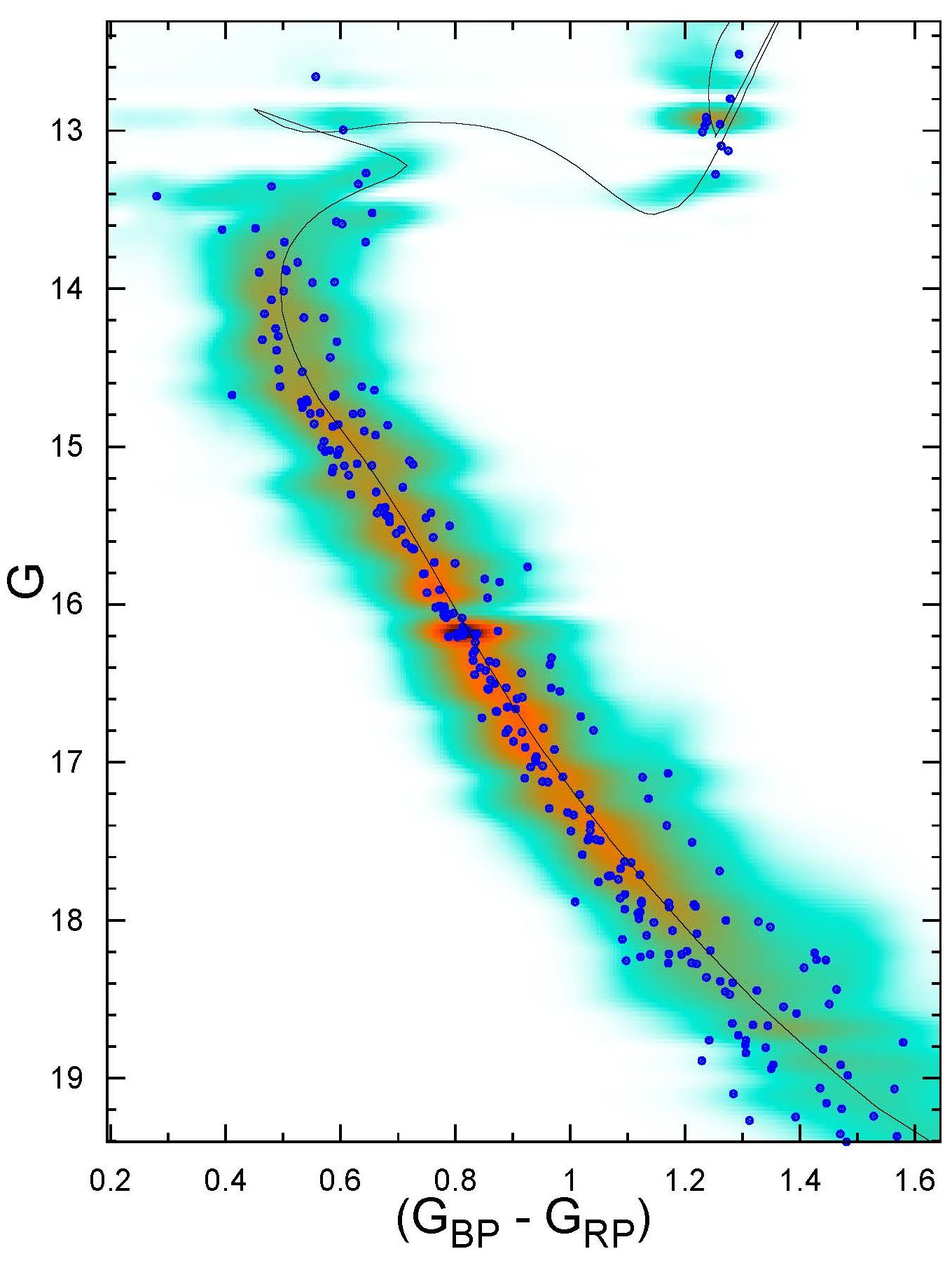}
	}
	\caption{$(G, G_{BP}-G_{RP})$ CMDs (Hess~diagram)  for Mel\,72 and Pismis\,7. The symbols and their meanings are the same as Fig.~4.}
\end{figure*}

\begin{figure*}[h!]\label{Fig-6}
	\centering{
		\includegraphics[width=0.24\textwidth]{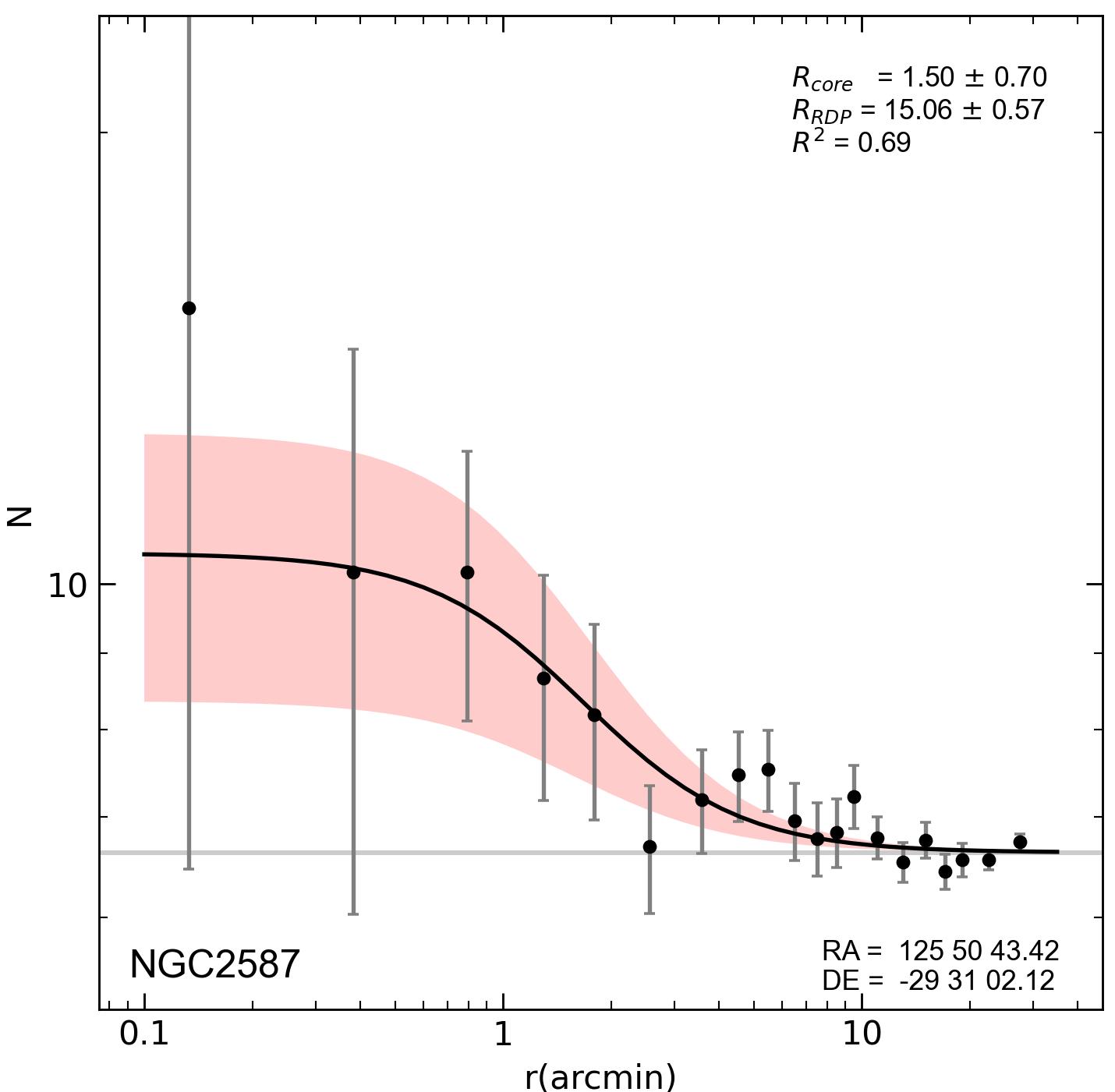}\hspace{2mm}
		\includegraphics[width=0.23\textwidth]{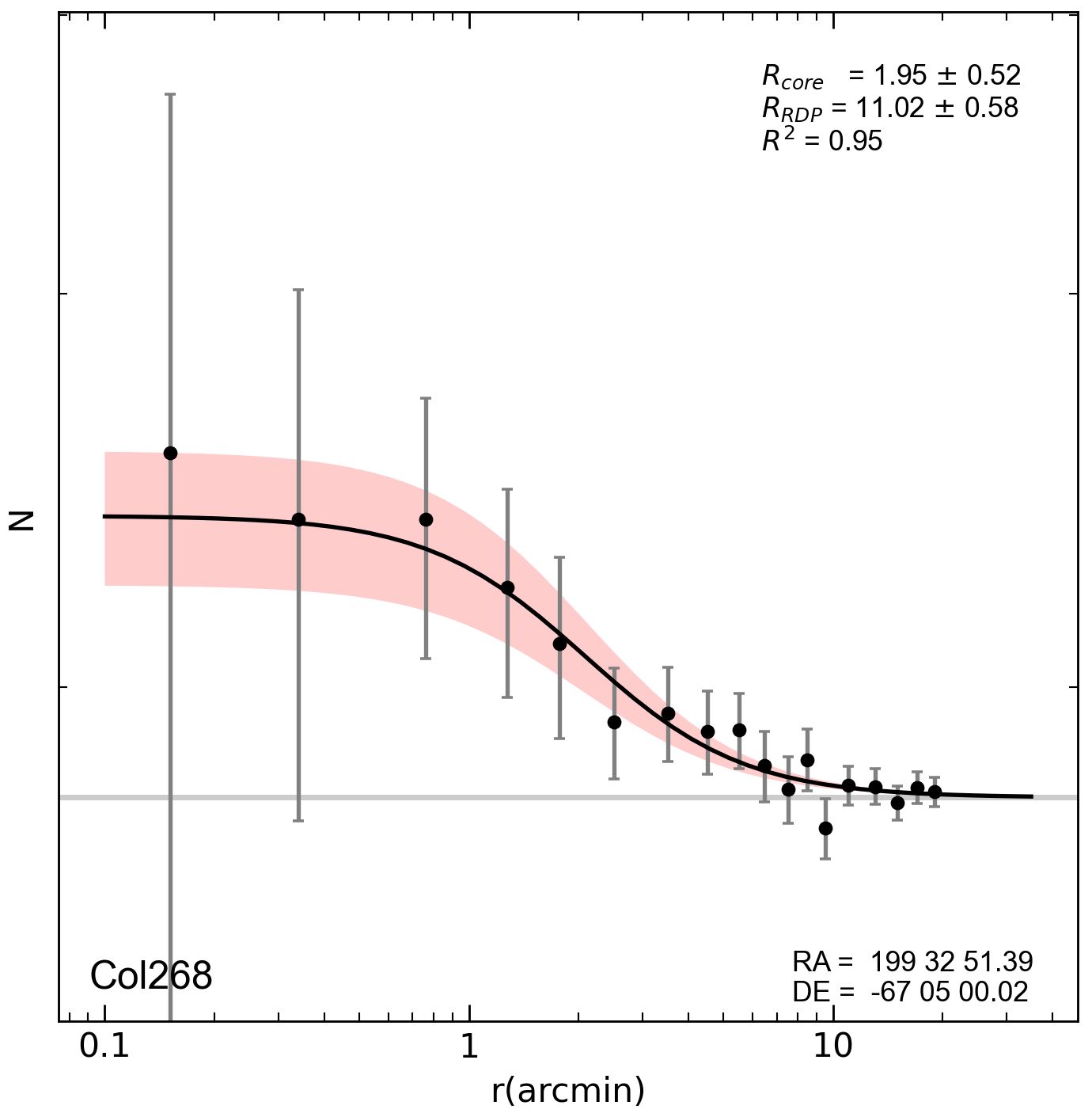}\hspace{2mm}
		\includegraphics[width=0.24\textwidth]{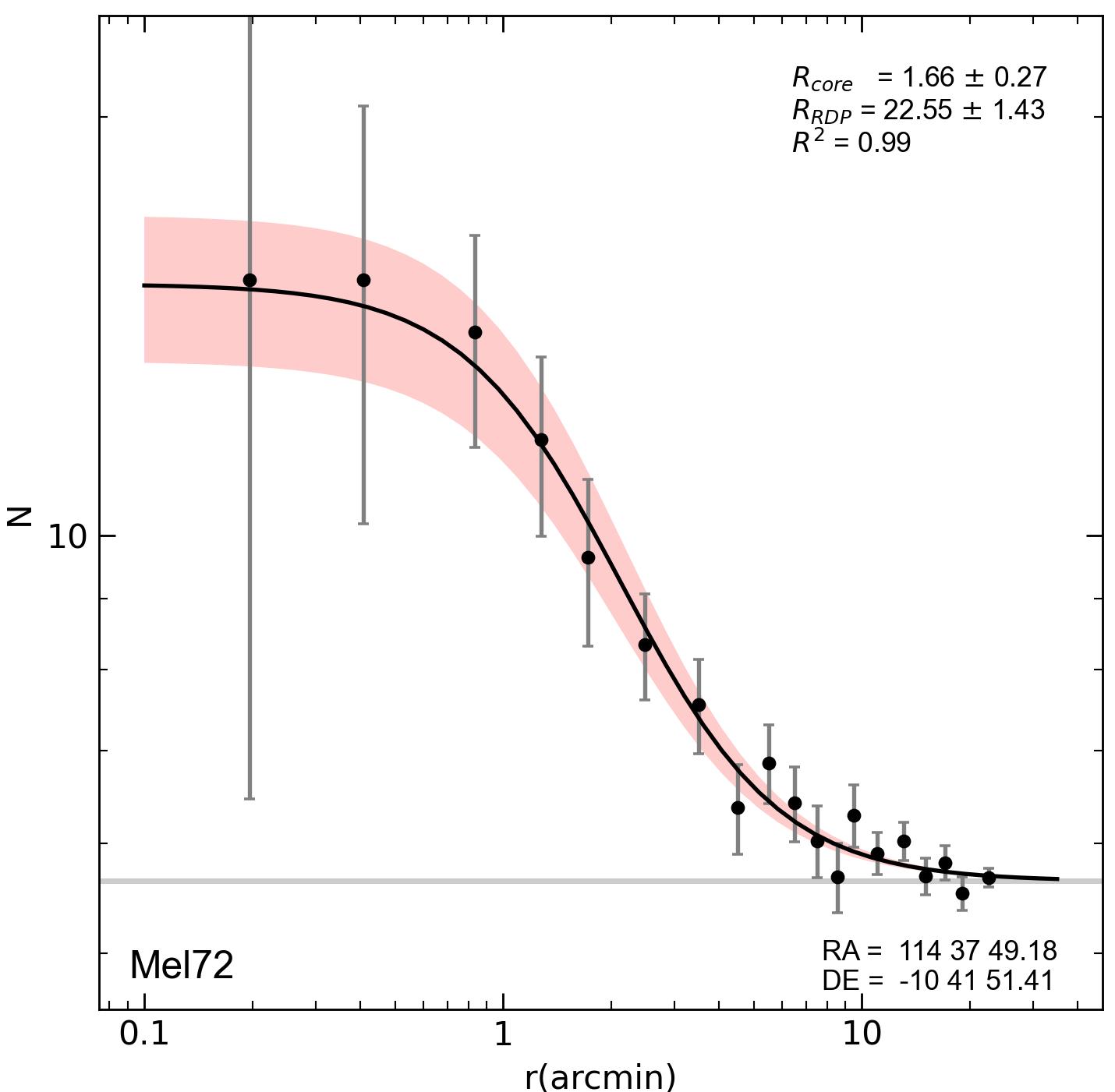}\hspace{2mm}
		\includegraphics[width=0.24\textwidth]{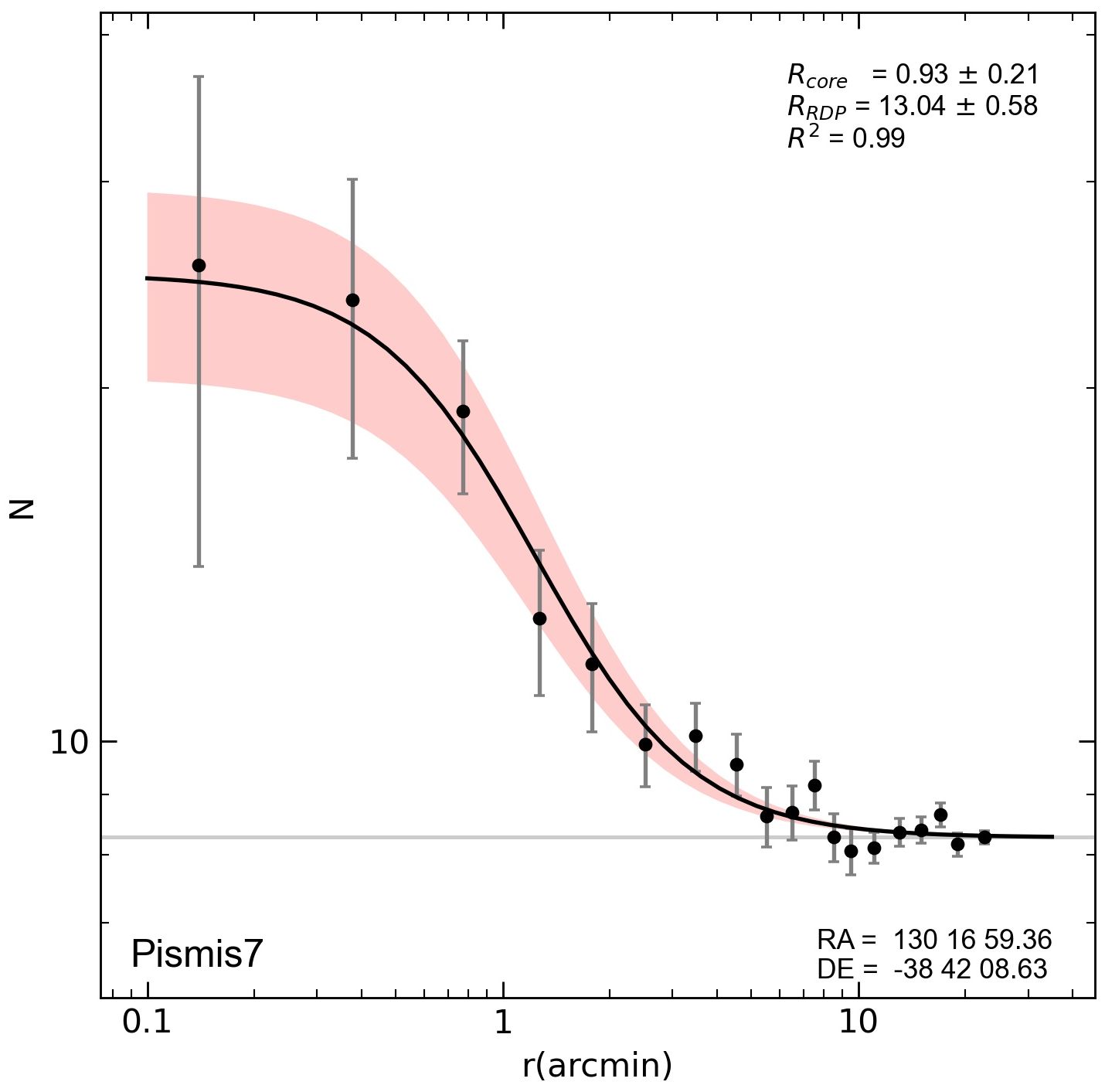}
	}
	\caption{Stellar RDPs (filled dots) of four OCs. Solid line shows the best-fit King profile. Horizontal bar: stellar background level measured in the comparison field. Shaded region: $1\sigma$ King fit uncertainty.}
\end{figure*}

\begin{figure}[t!]\label{Fig-7}
	\centering{\includegraphics[width=0.5\textwidth]{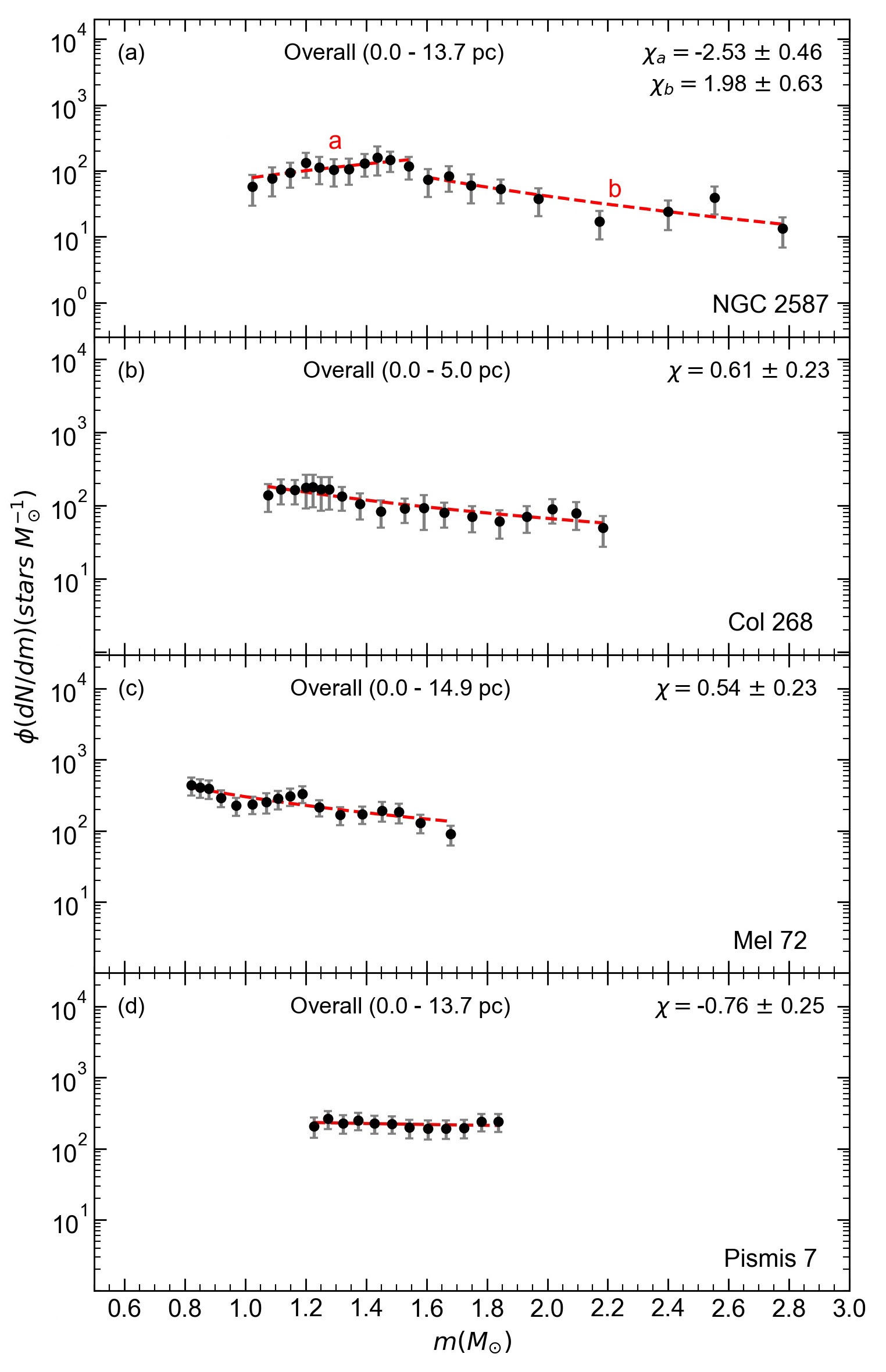}}
	\caption{$\phi(m)$ versus $m~(m_{\odot})$ for the overall regions of four OCs. 
		Here $\phi(m)= dN/dm$ (stars ~$m_\odot^{-1}$).}
\end{figure}

\section{Dimension and Mass Function}
By utilising the stellar radial density profiles (RDPs), we have derived structural parameters of four OCs. The fitted RDPs of four OCs to the relation of \cite{King1966} have been displayed in Fig.~6. For this, the three-parameter function, $\sigma(R) = \sigma_{bg} + \sigma_0/(1+(R/R_{core})^2$, given by \cite{King1966} is considered. For this equation, $\sigma_{bg}$ is the residual background density. $\sigma_0$ and $R_{core}$ are the central density of stars and the core radius, respectively.  
The meanings of solid line, horizontal bar and the shaded area are mentioned in caption of Fig.~6. We obtained the cluster radii $(R_{RDP})$ by comparing the RDP level with background and by measuring the distance from the centre. The $R_{core}$, $\sigma_{bg}$, and  $\sigma_0$ of four OCs have been derived from the fitted King profile to the observational RDPs (Fig.~6). The large uncertainties within  $R<1'$ in the RDPs are due to their low star contents in their central parts. These structural parameters are listed in Cols.~3-10 of Table~7. 

From the obtained masses from \textit{fitCMD} (Sect.~3), the mass function relation of $\phi(m)(stars ~m_\odot^{-1})$ versus m($m_\odot$) is displayed in Fig.~7 for the overall parts of four OCs. Table 4 provides the mass information. The observed number of stars and the corresponding total mass (obviously, the mass is computed by the simulation) are listed in Cols.~4-5. The mass range (Col~3) refers to the mass which is actually present in the observed CMDs, while $m_{tot}(m_\odot)$ and star number (Cols.~6-7) refer to \textit{fitCMD}'s full simulation (including stars with $>0.08~m_\odot$).  Ideally, the simulated number of stars should be exactly the same as the observed, but sometimes there are residual field stars. The full simulation corresponds to the whole mass range assumed to be present in the cluster. For instance, in the case of NGC~2587, the observed CMD contains 118 stars with $m_{tot}=180~m_\odot$ in the mass range: 0.96--3.09~$m_\odot$; if we had access to the full range, the stars would have masses between 0.08~$m_\odot$ to 3.09~$m_\odot$. Therefore, \textit{fitCMD} computes the values for this full range as 1621 stars storing 681~$m_\odot$.

The MF slopes shown in the plots are the observed ones, i.e., those computed directly from the observed CMDs. The theoretical IMF is used only for the purpose of estimating the completeness-corrected mass (by estimating the difference in the number of stars actually detected at a given magnitude with respect to the expected one). Therefore, all the other parameters are unaffected by this procedure.

The MFs of NGC 2587 (Fig.~7) presents a break followed by slope flattening for masses in the range $0.96 \leq m_\odot \leq 3.09$, which is noticeable particularly in the overall region. Up to $m \sim 1.5m_{\odot}$, its overall MFs is very steep negative $(\chi=-2.60\pm0.46)$.  After $m > 1.5m_{\odot}$, this MFs becomes very steep positive $(\chi=+1.98\pm0.63)$. Since the break occurs at $m \sim 1.5m_{\odot}$, it is not the same as that implied by Kroupa's mass function $(m \sim0.5m_{\odot})$. This break is due to completeness affecting masses lower than $m \sim 1.5m_{\odot}$. We use the MFs for  $m > 1.5m_{\odot}$ as representative of the overall cluster. This mass range does not seem to be critically affected by completeness.  The overall MF slopes of Col\,268 and Mel\,72 are quite positive/flat $(\chi=+0.61\pm0.23)$ and $(\chi=+0.54\pm0.23)$, respectively.  
Pismis\,7's overall MFs is negative/flat $(\chi=-0.76\pm0.25)$.

\renewcommand{\tabcolsep}{1.4mm}
\renewcommand{\arraystretch}{1.3}
\begin{table}[t!]\label{table-4}
	{\footnotesize
		\begin{center}
			\caption{Mass information for the overall regions of four OCs. 
			Their meanings are explained in Sect.4.}	
			\begin{tabular}{llcrccc}
				\hline
				Cluster& r~(pc) &mass range-$m(m_{\odot})$ &  N & $m_{tot}(m_{\odot})$ &  N & $m_{tot}(m_{\odot})$ \\
				\hline
				NGC~2587  & 0.0--13.7 &0.96-3.09 & 118 & 180 & 1621 & 681 \\
				Col.~268  & 0.0--5.0  &0.84-2.52 & 128 & 170 & 1901 & 748 \\
				Mel.~72   & 0.0--14.9 &0.66-2.06 & 273 & 310 & 2546 & 950 \\
				Pismis.~7 & 0.0--13.7 &1.06-2.18 & 204 & 230 & 2575 & 973 \\
				\hline
			\end{tabular}
		\end{center}
	}
\end{table}

\section{Relaxation time and Evolutionary parameter}
The relaxation times, $t_{rlx}$~(Myr) of four OCs are obtained from a relation, $t_{rlx}\approx0.04\left(\frac{N}{ln N} \right)\left(\frac{R}{1pc}\right)$. Here N is the number of stars located inside the cluster radius, $R_{RDP}$. As an indicator of dynamical evolution, the evolutionary parameters are estimated from the relation of $\tau = Age/t_{rlx}$. The relations of these time scales can be found in the works of \cite{Bonatto2005}, \cite{Bonatto2006} and \cite{Bonatto2007}. By adopting $\sigma_{v}\approx1\,km s^{-1}$ as an upper value explicitly \citep{Bonatto2011}, instead of $\sigma_{v}\approx3\,km s^{-1}$ of \cite{Binney1998}, these time scales are estimated and listed in Table~5.

$t_{rlx}$ gives the time required for the stars in the core/halo to travel from one end of these regions to the other (Stars move at $\sigma_{v}\approx1\,km s^{-1}$).  Mass segregation (known as migrating from the cores to halo) is directly related to $t_{rlx}$ and  $\tau$. $\tau$ also depends on both age and $t_{rlx}$. There is a negative relationship between $t_{rlx}$ and $\tau$. Large $\tau$ and small $t_{rlx}$ correspond to advanced mass segregation, accordingly, small $\tau$ and large $t_{rlx}$ mean small scale mass segregation.

From the propagating the errors in Age (Col.~5 of Table~6), radii and N (Col.6 of Table~4) into $t_{rlx}$ and $\tau$, the uncertainties of the evolutionary parameters $(\tau)$ are estimated. The errors in ages and radii are at a similar ($\sim3\%-13\%$) level. If the uncertainties in the number of stars (N) become to be larger, this is responsible for a large uncertainty in $t_{rlx}$ (Table~5) and, consequently, a large uncertainty in the evolutionary parameter. Here, $t_{rlx}$ and $\tau$ are considered simply as an order of magnitude estimate.

\renewcommand{\tabcolsep}{6.5mm}
\renewcommand{\arraystretch}{1.2}
\begin{table}[h!]\label{table-5}
	\centering
	\caption{Relaxation times and evolutionary parameters of overall regions of four OCs.}
		\begin{tabular}{lcc}
			\hline
		Cluster  & $t_{relax}\,(Myr)$ & $\tau$\\
		\hline
			NGC\,2587  & 1120.7$\pm$814.0 &  0.4$\pm$0.3 \\
			Col\,268   & 250.0$\pm$142.5  &  2.0$\pm$1.2 \\
			Mel\,72    & 592.7$\pm$341.0  &  2.1$\pm$1.2 \\
			Pismis\,7  &148.8 $\pm$105.0  &  6.7$\pm$4.8 \\
			\hline
		\end{tabular}
\end{table}

\renewcommand{\tabcolsep}{1.8mm}
\renewcommand{\arraystretch}{1.4}
\begin{table*}[t!]\label{table-6}
	\centering
	{\footnotesize
		\caption{Comparison of the astrophysical parameters to the literature. Cols.~1-4 represent the cluster name, reddenings, the metal/heavy element abundances, true distance moduli $(V-M_{V})_{0}$, and their corresponding heliocentric distances, respectively. Col.~5 gives the ages $(\log(A)/A~(Gyr))$. The isochrones, photometry and the references are listed in Cols.~6-8, respectively.}
		
		\begin{tabular}{lBBBBllc}
			\hline
			Cluster &\mcl{$E(B-V)$~~\&~~$E(G_{BP}-G_{RP})$} &  \mcl{[M/H]$~~\&~~$ Z} & \mcl{$(V-M_{V})_{0}$~~\&~~$d$~(pc)} 
			& \mcl{$\log(A) \& A~(Gyr)$} & \mcc{Isochrone}. & \mcc{Phot.} &  Reference \\
			\hline
			NGC\,2587&0.11&0.13& 0.22&0.025&12.48&3128& 8.65&0.45& Bressan et al.(2012) & Gaia~DR2 &This paper \\
			& 0.23&-     &  \mcl{-}    & 11.20&1740 & 8.00&0.10 & Bonatto et al.(2004)     & 2MASS      &5\\
			& 0.10&-     &  0.02&0.02  & 12.18&2700 & 8.70&0.50 & Lejuene\&Schaerer (2001) & UBVI       &6\\
			& 0.09&-     & \mcl{solar} & 12.55&3250 & 8.60&0.40 & Girardi et al.(2002)     & 2MASS      &3\\
			& 0.10&-     & \mcl{solar} & 12.19&2700 & 8.70&0.50 & Marigo et al.(2008)      & 2MASS      &4\\
			&0.15~(A_{V}=0.45)&- & \mcl{solar} & 12.53&3210 & 8.50&0.32 & Bressan et al.(2012)  & Gaia~DR2 &10\\
			\hline
			Col\,268 &0.53&0.66&-0.78&0.0025&10.97&1564&8.70&0.50& Bressan et al.(2012)& Gaia~DR2  &This paper \\
			& 0.36&- &  \mcl{-}    & 11.23&1740 &  \mcl{-}  &  -                  & $ubvyH_{\beta}$ &8\\
			& 0.23&- & \mcl{solar} &     -&1900 & 8.95&0.60 & Girardi et al.(2002)  & 2MASS     &9\\
			& 0.24&- & \mcl{solar} & 11.10&1680 & 8.85&0.71 & Girardi et al.(2002)  & 2MASS     &3\\
			& 0.34&- & \mcl{solar} & 11.40&1810 & 8.79&0.62 & Marigo et al.(2008)   & 2MASS     &4\\
			&0.37~(A_{V}=1.16)&- & \mcl{solar} & 12.10&2630& 8.38&0.24& Bressan et al.(2012)  & Gaia~DR2 &10\\
			\hline
			Mel\,72 &0.16&0.19&-0.14&0.011&11.79&2277& 9.10&1.25& Bressan et al.(2012)& Gaia~DR2&This paper \\
			& 0.20&-   & \mcl{solar} & 12.38&3000 & 8.80&0.60 & MAI                   & $UBVI_{KC}$ &1\\
			& 0.08&-   & -0.37&0.008 & 12.51&3180 & 9.20&1.60 & Bertelli et al.(2004) & BVI         &2\\
			& 0.07&-   & \mcl{solar} & 12.03&2550 & 8.95&0.89 & Girardi et al.(2002)  & 2MASS       &3\\
			& 0.23&-   & \mcl{solar} & 13.06&3960 & 8.86&0.72 & Marigo et al.(2008)   & 2MASS       &4\\
			& 0.14&0.18& \mcl{solar} & 12.22&2340 & 9.00&1.00 & Marigo et al.(2017)   & Gaia~DR2    &11\\
			&0.13~(A_{V}=0.41)&- & \mcl{solar} & 12.14&2680& 8.99&0.98& Bressan et al.(2012)  & Gaia~DR2 &10\\
			\hline
			Pismis\,7 &0.70&0.86&-0.28&0.008&12.79&3614& 9.00&1.00& Bressan et al.(2012)& Gaia~DR2&This paper \\
			& 0.69&-  & \mcl{solar} & 13.46&4920 & 8.70&0.50 & Girardi et al.(2000)  & BVRI      &7\\
			& 0.36&-  & \mcl{solar} & 14.07&6690 & 8.90&0.79 & Girardi et al.(2002)  & 2MASS     &3\\
			& 0.94&-  & \mcl{solar} & 13.70&4780 & 8.70&0.50 & Marigo et al.(2008)   & 2MASS     &4\\
			&0.52~(A_{V}=1.60)&- & \mcl{solar} & 13.32&4610& 8.89&0.78& Bressan et al.(2012)  & Gaia~DR2 &10\\
			\hline
		\end{tabular}
		\\
		\begin{list}{Table Notes.}
			\item (1):Piatti et al. (2010), (2):Hasegawa et al. (2008); (3): Bukowiecki et al.(2011); (4): Kharchenko et al. (2013); (5): Tadross (2011); (6):Piatti et al. (2009); (7): Ahumada (2005);(8): Moffat et al. (1975), (9) :Bica et al. (2008), (10): Cantat-Gaudin et al. (2020). (11): Hendy and Tadross (2021). MAI~(Col.~6) means the morphologic age index.
		\end{list}
	}
	
\end{table*} 

\renewcommand{\tabcolsep}{1.3mm}
\renewcommand{\arraystretch}{1.4}
\begin{table*}[t!]\label{table-7}
	{\footnotesize
		\begin{center}
			\caption{Structural parameters and literature comparison of four OCs. Col. 2: arcmin to parsec scale. The symbol $*\,\prime^{-2}$ in Cols.~7-8 mean $stars~arcmin^{-2}$. Comparison field ring and the correlation coefficient~(CC) are listed in Cols.~9-10. Total mass, mass function slope~(MFs), total star number, references, respectively are given in Cols.~11-14.} 
			\begin{tabular}{lcccccccccccccc}
				\hline
				Cluster & $(1')~(pc)$  & $R_{core}$& $R_{RDP}$ & $R_{core}$& $R_{RDP}$ & $\sigma_{0K}$ & $\sigma_{bg}$ & $\Delta R$ &CC&$m_{tot}(m_{\odot})$&MFs&N& Reference\\
				&  &$pc$ & $pc$  & $\prime$& $\prime$& $*\,\prime^{-2}$&$*\,\prime^{-2}$& $\prime$& & \\
				$1$ & $2$ & $3$ & $4$ & $5$ & $6$ & $7$ & $8$& $9$ &$10$&$11$&$12$&$13$&$14$\\
				\hline
				NGC\,2587&0.91&1.40$\pm$0.60&13.70$\pm$0.50&1.5$\pm$0.7&15.10$\pm$0.60&3.9$\pm$2.1&6.63$\pm$0.05&16-25&0.69&681&1.98$\pm$0.63&1621&This paper\\
				&    &0.69$\pm$0.10&5.53$\pm$0.79&$-$ &$-$ &$-$ &$-$ &$-$ &$-$ &174&-1.13&105$\pm$135&1\\
				&    &1.71$\pm$0.49& $-$&$-$ &$-$ &$-$ &$-$ &$-$ &$-$ &$-$ &$-$&$-$&2\\
				\hline
				Col\,268 &0.45&0.90$\pm$0.2&5.0$\pm$0.30&2.0$\pm$0.5&11.0$\pm$0.6&6.0$\pm$1.6&17.84$\pm$0.09&12-31&0.95&748&0.61$\pm$0.23&1901&This paper\\
				&    &0.97$\pm$0.14&5.62$\pm$0.72&$-$ &$-$ &$-$ &$-$ &$-$ &$-$ &478&2.05&1369$\pm$1230&1\\
				&    &0.75$\pm$0.27& $-$&$-$ &$-$ &$-$ &$-$ &$-$ &$-$ &$-$ &$-$&$-$&2\\
				&    &0.60$\pm$0.20&3.70$\pm$0.30 &$-$ &$-$ &$-$ &$-$ &$-$ &$-$ &$-$ &$-$&$-$&3\\
				\hline
				Mel\,72  &0.66&1.10$\pm$0.20&15.00$\pm$0.90 & 1.7$\pm$0.3&22.6$\pm$1.4 &9.5$\pm$1.3 &5.63$\pm$0.05&23-31& 0.99&950&0.54$\pm$0.23&2546&This paper\\
				&    &0.80$\pm$0.10&8.2$\pm$1.00&$-$ &$-$ &$-$ &$-$ &$-$ &$-$ &647&0.70&7279$\pm$616&1\\
				&    &2.21$\pm$0.38& $-$&$-$ &$-$ &$-$ &$-$ &$-$ &$-$ &$-$ &$-$&$-$&2\\
				&    &0.31$\pm$0.06& 3.41$\pm$0.10 &0.45$\pm$0.09 &5.0$\pm$0.15 &$-$ &$-$ &$-$ &$-$&223&-2.20$\pm$0.10&168&4\\
				\hline
				Pismis\,7&1.05&0.90$\pm$0.20&13.70$\pm$0.60& 0.9$\pm$0.2 & 13.0$\pm$0.6&16.7$\pm$4.6&8.27$\pm$0.06&14-21&0.99&973&-0.76$\pm$0.25&2575&This paper\\
				&    &1.50$\pm$0.18&14.05$\pm$2.05&$-$ &$-$ &$-$ &$-$ &$-$ &$-$ &$-$ &$-$&$-$&1\\
				&    &2.10$\pm$0.95& $-$&$-$ &$-$ &$-$ &$-$ &$-$ &$-$ &$-$ &$-$&$-$&2\\
				\hline
			\end{tabular}
			\begin{list}{Table Notes.}
				\item (1): Bukowiecki et al.(2011); (2): Kharchenko et al. (2013); (3):Bica et al. (2008); (4): Hendey and Tadross (2021).
			\end{list}
		\end{center}
	}
\end{table*}

\begin{figure}[h!]\label{Fig-8}
	\centering{\includegraphics[width=0.85\columnwidth]{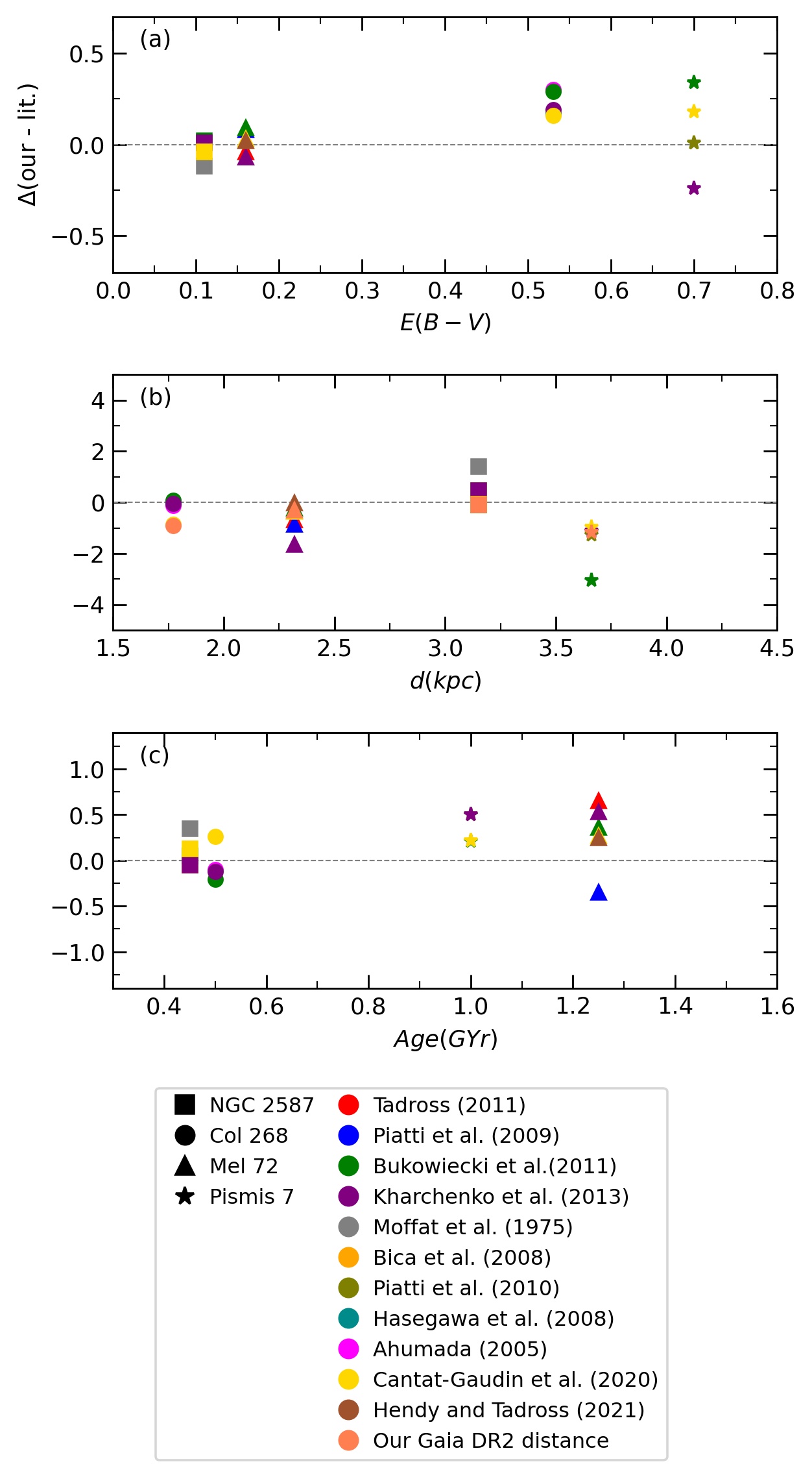}}
	\caption{The differences between this paper and literature for $E(B-V)$, $d~(kpc)$, and $Age$~(Gyr) in Table~6.}
\end{figure}

\begin{figure}[h!]\label{Fig-9}
	\centering{\includegraphics[width=0.95\columnwidth]{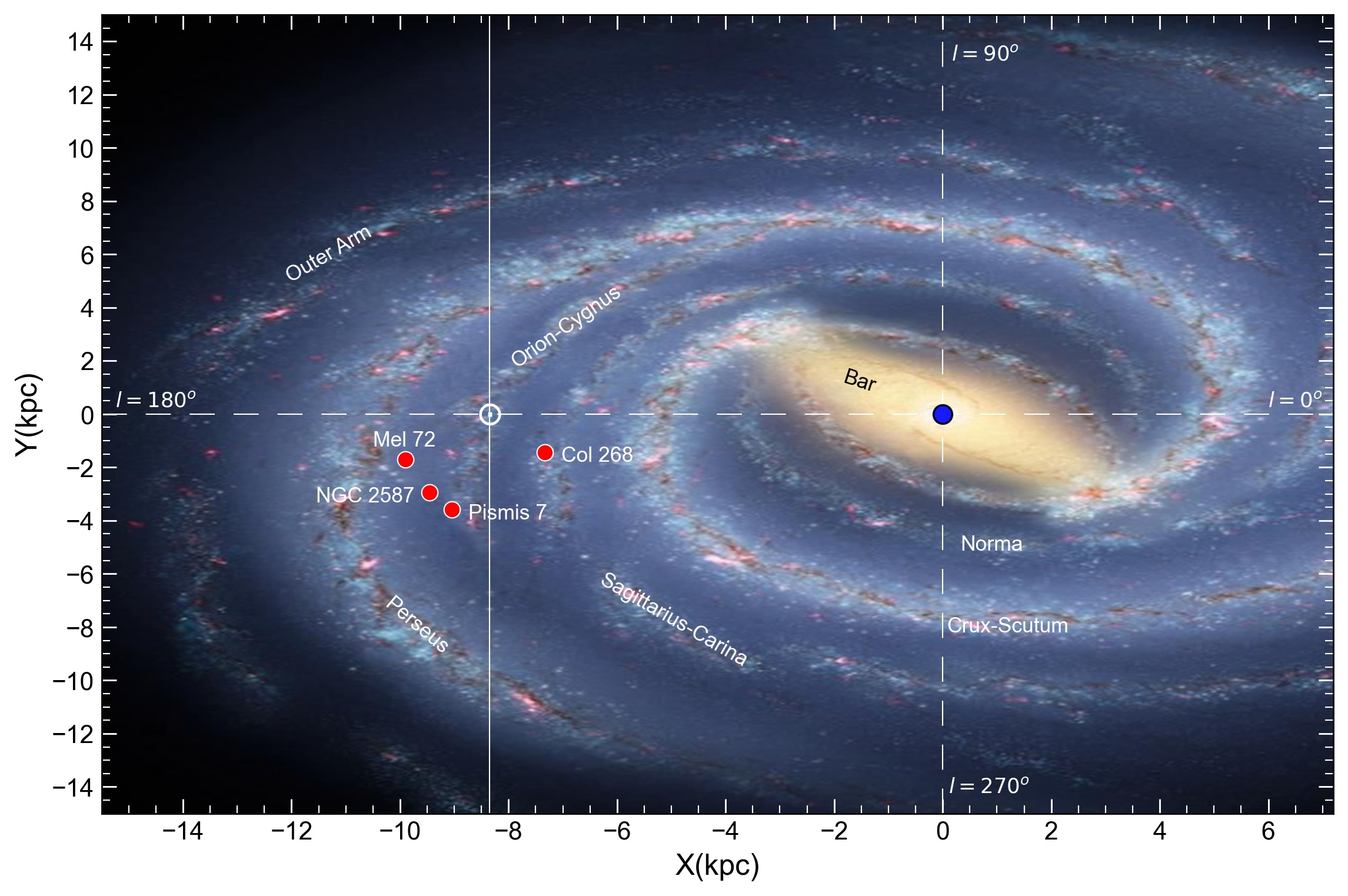}}
	\caption{Spatial distribution of our sample OCs (filled circles). 
		The schematic projection of the Galaxy is seen from the North pole. $(X_{GC},~Y_{GC})$~kpc show the Galactocentric cartesian coordinates.}
\end{figure}
 
\begin{figure}[h!]\label{Fig-10}
	\centering{\includegraphics[width=0.95\columnwidth]{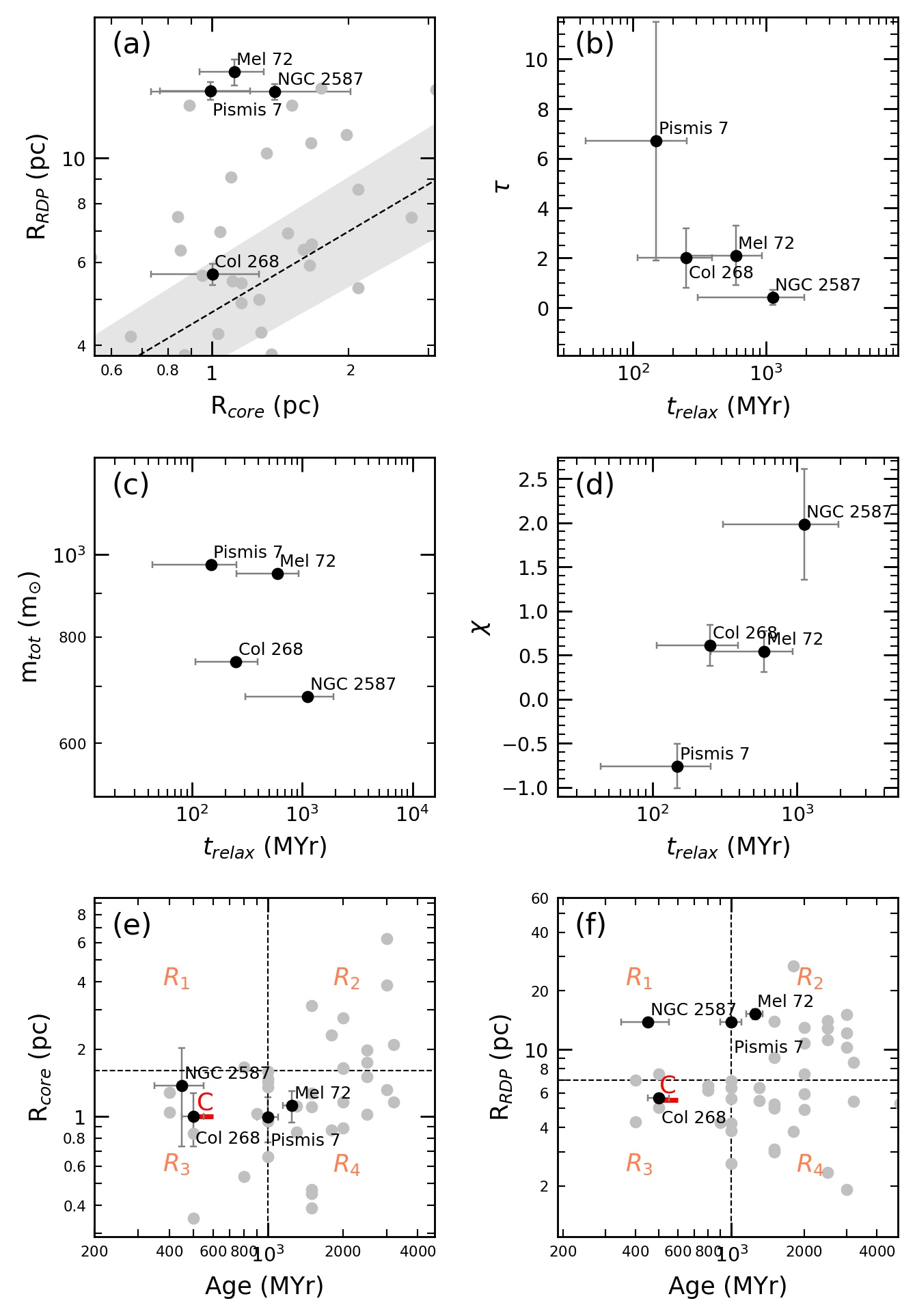}}
	\caption{For four OCs, $R_{RDP}$ versus $R_{core}$  (panel~a),  $\tau$ versus $t_{rlx}$ (panel~b), $m_{tot}(m_\odot)$ versus $t_{rlx}$ (panel c), $\chi$ versus $t_{rlx}$ (panel~d),  $(R_{core},~R_{RDP})$ versus Age (panels~e--f). The meanings of the symbols in panels (a), (e), (f) are explained in the text.}
\end{figure}	

\section{Discussion and Conclusion}
\subsection{Comparison with the literature}
The astrophysical parameters of four OCs have been determined from \textit{fitCMD} algorithm. A comparison of $E(B-V)$, $E(G_{BP}-G_{RP})$,~$Z$,~$(V-M_{V})_{0}$, ~$d~(pc)$ and $Age$~(Gyr) values of four OCs with the literature is presented in In Table~6/Fig.~8. The reddening comparison has been given in $E(B-V)$, which is converted from Gaia/2MASS.  The small/large differences between our values and the literature can be seen from Table~6/Fig.~8.  We leave the detail comparison to the reader.

Reddenings and the photometric distances which are found from \textit{fitCMD} for NGC\,2587 and Mel\,72 are in good agreement with the ones of \cite{cantat2020}.  Differences of the reddenings and distances for Col\,268 and Pismis\,7 are up to 0.16--0.18 mag  and  $\sim1.0$~ kpc, respectively. This stems from the usage of $Z$ values of \textit{fitCMD}, instead of solar abundance. 
For Mel\,72, our reddenings/distances/ages are in good agreement with \cite{Hendey2021}. 
The obtained photometric distances from \textit{fitCMD} provide somewhat close distances, as compared to  Gaia DR2 distances (Col.6 of Table 2). For NGC\,2587 and  Mel\,72, both distances are in good concordance. The differences of both distances are at a level of $\sim1.0$~ kpc for Col\,268 and Pismis\,7.

The results for Pismis 7 are neccessarily rather tentative. It is the most distant and the most
heavily reddened cluster and clearly shows that its stars are differentially reddened.

As emphasized by \cite{Paunzen2006} and \cite{Moitinho2010}, the small/large discrepancies in ages and distance moduli/distances (Table~6) result from the adopted isochrones and the reddening values, which are found from CMDs.  Although the B12 Padova isochrones are used for this paper and the papers of B11 and K13, large/small differences in ages can be explained by the derived reddenings. 

The dimensions, total masses, MF slopes and star numbers of four OCs  are presented in Table~7 together with the literature values. The inconsistency of these parameters with the literature may be partly explained by the detected star numbers (Col.~13 of Table 7).

\subsection{Dynamical Evolution of Four OCs}
The relations between dynamical evolution parameters are presented  in Figs.~10(a)--(f). The relations, the horizontal lines, the labels $R1$-$R4$, "C" and filled grey points of panels (a), (e), (f) are from \cite{Gunes2017} (their figs.~ 11, 14, and 17). From panel~(a), NGC\,2587, Mel\,72 and Pismis\,7 do not follow the relation of \cite{Gunes2017}.
These OCs can be partly attributed to clusters with large radii retaining their masses.  Whereas Col\,268 with the small dimensions is quite close to the relation (dashed diagonal).
	
The relaxation times/evolutionary parameters of four OCs exhibit similarity because they are all within one sigma of one another (Table 5 and panel~b).
Except NGC 2587, the ages of the remaining OCs are higher than their relaxation times. So, they are dynamically relaxed. From panel c, they seem to be less massive OCs than $m_{tot}=1000 m_{\odot}$.
Mass function gives the  distribution of the stars in the cluster. In this context, NGC\,2587's steep MFs $(\chi=+1.98)$ means that its low-mass stars outnumber its massive ones (panel d).  This OC did not undergo dynamical evolution due to its small evolutionary time. This explains  the reason that this OC lies in $R1$/$R3$ with large cluster/small core radii (panels e and f).
The relatively flat (as compared to $\chi=+1.35$ of \cite{sal55} or $\chi=1.30$ of \cite{Kroupa2001}) MF slopes of Col\,268 and especially Mel\,72 (panel d) suggest that both OCs present signs of a somewhat advanced dynamical evolution, in the sense that they appear to have lost a significant fraction of their low-mass stars to the field.

Col\,268 with the small dimensions $(R_{core},~R_{RDP})=(0.9,~5.0)$~pc is intrinsically small.  Instead of shrinking in size and mass with time, it may have a primordial origin which may be related to high molecular gas density in Galactic directions \citep{Camargo2010, van91}. Note that Col\,268 (0.5 Gyr) falls in the range "C" (panels~e--f).

Pismis\,7's negative/flat MFs (panel d) implies that its high-mass stars slightly outnumber its low-mass stars. Due to its $\tau=6.7$ value, Pismis\,7 shows a sign of mild dynamical evolution. In this context, its high mass stars move towards the central region, while its low mass stars are continually being lost to the field. Regarding Pismis\,7, it is the most distant OC, so completeness may affect the detection of its low-mass content. However, since its age is about 1 Gyr, mass segregation also may played some role in depleting its low-mass stars. The combination of these two factors may explain its MFs.
 
In panels~e--f, the cluster dimension versus cluster age is given. As discussed by \cite{Camargo2010}, this kind of relationship provides some information for cluster survival/dissociation rates. From this perspective, some clusters expand with time, while some seem to shrink.
According to \cite{Mackey2003}, the dynamical evolution of the core/cluster radii of the clusters started at 500-600 Myr (the label "C", panels~e--f), and continued to 1 Gyr. A bifurcation occurs at an age $\sim 1$ Gyr.  Pismis\,7 locates at the bifurcation. Mel\,72  locates in $R2$ region.  Therefore, the outer parts of Mel\,72 and Pismis\,7 expand with time. However,  Mel\,72' core contracts because of dynamical relaxation, dependig on its $R4$ location.  However, Pismis\,7's core may expand with the time because its massive stars move towards its central parts, given its mild dynamical evolution.

Considering the locations of four OCs in the Galaxy (Table~1/Fig.~9), NGC\,2587, Mel\,72 and Pismis\,7 locate at third quadrant (outside solar circle), which is a region with low density of GMCs. So, except for Pismis\,7,  they do not seem to expose to the external dynamical effects much, such as tidal stripping due to disk and bulge crossings plus encounters with GMCs.  Col\,268 did not seem to lost its star content much because of the presence of massive GMCs, and tidal effects from disk and Bulge crossings as external process, taking care its direction $(\ell=305^{\circ}.54,~R_{GC}=7.6~kpc)$\footnote{We adopt the value $R_{\odot}=8.2\pm0.1$ kpc of \cite{Bland-Hawthorn2016} for their $R_{GC}$ distances.}

Star clusters usually present crowding, especially towards the central region. This means that, near the center, some fraction of the faintest stars will not be detected because of crowding. At the outer parts where crowding is less important, one can get more faint stars observed. However, two factors should be considered on our results. A large uncertainty in $t_{rlx}$ and $\tau$ reduces the reliability of the interpretation of the dynamic evolution of these OCs. In this respect, for the interpretations we consider their mean values as a simple approach. Binaries widen the main sequence of the OCs by 0.75 mag, so the theoretical isochrones are fitted to the mid-points of CMDs of the OCs, rather than the faint or blue sides \citep{Carney2001}. Due to a consequence of the dynamical evolution in OCs, multiple systems tend to concentrate in central regions \citep{Takahasi2000}. Because of a significant fraction of binaries in the central parts of OCs, the number of low-mass stars is underestimated with respect to the high mass stars \citep{Bonatto2005}.

\section{Acknowledgments}
We thank our referee for his/her valuable suggetsions/comments. This paper has made use of results from the European Space Agency (ESA) space mission Gaia, the data from which were processed by the Gaia Data Processing and Analysis Consortium (DPAC). Funding for the DPAC has been provided by national institutions, in particular the institutions participating in the Gaia Multilateral Agreement. The Gaia mission website is http: //www.cosmos.esa.int/gaia. This paper has also made use of the WEBDA database, operated at the Institute for Astronomy of the University of Vienna. This publication also makes use of SIMBAD database-VizieR (http://vizier.u-strasbg.fr/viz-bin/VizieR?-source=II/246.).

\nocite{*}


\end{document}